\documentclass[preprint,12pt]{article}
	
\usepackage[T1]{fontenc}
\usepackage[utf8]{inputenc}
\usepackage[english]{babel}

\usepackage[a4paper,margin=30mm]{geometry}
\usepackage{lmodern}
\usepackage{microtype}
\usepackage{setspace}
\usepackage{amsmath, amssymb}

\usepackage{graphicx}
\usepackage{booktabs}
\usepackage{threeparttable}
\usepackage{siunitx}
\usepackage{subcaption}
\usepackage{amsmath}
\usepackage{enumitem}
\usepackage{float}
\usepackage[colorlinks=true,linkcolor=blue,citecolor=blue,urlcolor=blue]{hyperref}

\title{Characterizing Agent-Based Model Dynamics via $\varepsilon$-Machines and Kolmogorov-Style Complexity}

\author{Roberto Garrone\thanks{Department of Informatics, Systems and Communication (DISCo),
		University of Milan--Bicocca, Viale Sarca 336, Milan, Italy}\\[0.5ex]
	\small University of Milan--Bicocca, Milan, Italy\\
	\small Open University of Cyprus, Nicosia, Cyprus\\
	\small \texttt{roberto.garrone@unimib.it}
}

\begin{document}
\maketitle

\begin{abstract}
We present a two-level, information-theoretic approach for analyzing the temporal organization of outputs generated by agent-based models (ABMs), framed within the study of complex adaptive systems. At a system-wide level, a pooled $\varepsilon$-machine is reconstructed from simulated time series to provide a coarse baseline of overall predictability. At an agent level, $\varepsilon$-machines are estimated separately for each caregiver–elder dyad and model variable, and are paired with compression-based complexity measures derived from lossless coding and normalized LZ78 statistics.

These diagnostics support comparative and distributional analyses across agents, variables, and simulation scenarios. Applied to a simulated caregiving model, the results indicate that pooled dynamics are effectively memoryless under coarse representations, whereas dyad-level analyses recover localized temporal structure. In particular, spatial accessibility indicators exhibit bounded but nontrivial temporal organization when ordinal encodings are used, while socioeconomic variables primarily reflect cross-sectional heterogeneity with little short-range memory.

Compression-based measures are consistent with these findings: dictionary-based compressors yield concordant estimates of descriptive redundancy, while normalized LZ78 highlights differences in statistical novelty across variables. Taken together, the results illustrate how predictive structure and descriptive compressibility capture distinct aspects of simulated dynamics. The proposed approach is intended as a diagnostic framework for exploratory analysis of agent-based simulation outputs, rather than as a contribution to computational mechanics theory.
\end{abstract}





\section{Introduction: Complex Adaptive Systems and ABM}

Complex Adaptive Systems (CAS) provide a theoretical paradigm for describing decentralized, evolving systems driven by feedback and adaptation, while Agent-Based Modeling (ABM) operationalizes this paradigm as a computational experiment. Both perspectives emphasize emergence, heterogeneity, and nonlinearity as core drivers of complex social and ecological dynamics. Integrating CAS theory with ABM practice enables researchers to link micro-level rules with macro-level phenomena and to explore resilience, innovation, and systemic sustainability \textit{in silico} \cite{Holland1992,Bonabeau2002,EpsteinAxtell1996,MillerPage2007}.

A CAS can be conceptualized as a network of interacting and interdependent components whose local behaviors collectively generate emergent, system-level patterns that are not reducible to the attributes of individual elements. Within such systems, agents adapt to both environmental conditions and the evolving behavior of others, producing feedback loops that continually reshape the system's structure and dynamics \cite{Holland1992,Holland2006}. The hallmark properties of CAS---nonlinearity, feedback, adaptation, and emergence---explain why even simple local rules can yield diverse macroscopic outcomes \cite{Levin1998,MillerPage2007,Bonabeau2002}. As emphasized by Bonabeau~\cite{Bonabeau2002}, global structure ``emerges'' from localized rules rather than centralized design, creating recursive micro--macro feedback and path dependence \cite{Levin2002}.

ABM provides a computational framework for representing and analyzing the mechanisms underlying complex adaptive behavior. In ABM, a system is represented as a collection of autonomous, heterogeneous agents---each with states, goals, and behavioral rules---interacting within a shared environment \cite{EpsteinAxtell1996,Epstein1999}. Through repeated interactions, these agents produce emergent macro-level dynamics that can be analyzed to understand how collective behavior arises from individual decisions. ABM adopts a generative epistemology: it seeks to ``grow'' phenomena from micro-level mechanisms rather than impose top-down equations \cite{Epstein2012}. This makes it a natural complement to CAS theory, capable of exploring how adaptation, feedback, and heterogeneity shape system evolution \cite{MillerPage2007,Bonabeau2002}. Conceptually, CAS defines the phenomenon, while ABM provides the method for its exploration \cite{Epstein1999,Bonabeau2002,MillerPage2007}.

\subsection*{Scope and contributions}
Despite the widespread use of agent-based models, the analysis of their outputs
is still dominated by static summaries and low-order temporal diagnostics,
such as averages, distributions, and autocorrelation functions. These tools
capture variability but provide limited insight into how simulated agents
store, transmit, or reuse information over time, particularly at the individual
or dyadic level. As a result, there is a methodological gap between the rich
temporal dynamics generated by ABMs and the comparatively coarse diagnostics
typically used to interpret them.

This study is methodological in nature. It does not introduce new theoretical results in computational mechanics, nor does it aim to empirically validate agent-based models against observed social data. Instead, it addresses a practical and under-formalized problem in the analysis of agent-based simulations: how to diagnose and compare the temporal organization of simulated agent-level dynamics beyond standard summary statistics.

The paper makes three specific contributions. First, it operationalizes established tools from computational mechanics and information theory—namely $\varepsilon$-machine reconstruction and compression-based complexity measures—as diagnostics for time series generated by agent-based models. Second, it clarifies the distinct roles of predictive structure and descriptive compressibility by jointly analyzing $\varepsilon$-machine–based measures and algorithmic compression proxies, showing how these capture complementary aspects of simulated dynamics. Third, it demonstrates the approach on outputs from a caregiver–elder agent-based model, illustrating how spatially mediated variables exhibit bounded temporal organization at the agent level, while socioeconomic variables primarily reflect cross-sectional heterogeneity.

All results are conditional on the simulation design and representation choices adopted. The proposed framework is intended as an exploratory and comparative diagnostic tool for agent-based modeling studies, rather than as a contribution to the theory of complex adaptive systems or computational mechanics.

\section{Computational Mechanics for ABM: $\varepsilon$-Machines}

The application of $\varepsilon$-machine reconstruction to Agent-Based Modeling
(ABM) remains uncommon, yet it is conceptually well grounded within the theory of
information processing in complex systems. Originally developed in nonlinear and
symbolic dynamics~\cite{CrutchfieldYoung1989,Crutchfield1994}, the
$\varepsilon$-machine provides the minimal unifilar model that partitions past
observation histories into sets---called \emph{causal states}---that yield
identical conditional distributions over possible futures~\cite{ShaliziCrutchfield2001}.
Formally, if two pasts $x_{:t}$ and $x'_{:t}$ satisfy
\begin{equation}
	\mathbb{P}\!\big(X_{t:\infty} \mid X_{:t}=x_{:t}\big)
	= \mathbb{P}\!\big(X_{t:\infty} \mid X_{:t}=x'_{:t}\big),
	\label{eq:causal_state_equiv}
\end{equation}
they belong to the same causal state. The $\varepsilon$-machine thus defines the
minimal predictive architecture of the process, specifying how information from
the past is stored and used to generate the future.

Three canonical quantities summarize its informational organization: the
\emph{entropy rate} $h_{\mu}$ (average unpredictability per symbol), the
\emph{statistical complexity} $C_{\mu}$ (information stored in the causal-state
distribution), and the \emph{excess entropy} $E$ (shared information between past
and future). In concise form:
\begin{equation}
	h_{\mu} = H[X_t \mid X_{:t}],
	\label{eq:entropy_rate}
\end{equation}
\begin{equation}
	C_{\mu} = H[\mathcal{S}],
	\label{eq:statistical_complexity}
\end{equation}
\begin{equation}
	E = I[X_{:t};\, X_{t:\infty}],
	\label{eq:excess_entropy}
\end{equation}
where $H[\cdot]$ and $I[\cdot;\cdot]$ denote Shannon entropy and mutual
information, respectively, and $\mathcal{S}$ is the causal-state random variable.
These quantities capture how much information a process generates, stores, and
transmits over time, thereby distinguishing structured dynamics from mere
randomness.

In the context of ABMs, $\varepsilon$-machines provide a principled means to
extract and quantify the system’s intrinsic computation—the transformation of
past information into future behavior. Here, causal states may correspond to
recurrent behavioral regimes, coordination cycles, or macro-level feedback
patterns that emerge from agent interactions. Unlike scalar measures such as
entropy or mutual information alone, the $\varepsilon$-machine reconstructs the
minimal predictive model: it reveals the architecture of information storage and
flow that governs the system’s evolution. This makes it particularly suited to
studying adaptive or path-dependent ABM dynamics, where emergent regularities
coexist with stochastic fluctuations.

In contrast, Kolmogorov--Solomonoff--Chaitin complexity formalizes the
descriptional compressibility of a sequence as the length of the shortest program
that can reproduce it on a universal Turing machine~\cite{LiVitanyi2008}. Because
true Kolmogorov complexity is non-computable, compression-based estimators such as
Lempel--Ziv provide practical proxies~\cite{CoverThomas2006}. The achieved
compressed size reflects the regularity and redundancy of the sequence. A general
compression-based complexity proxy is defined as
\begin{equation}
	C_{\mathrm{comp}} = \frac{8\, S_{\mathrm{comp}}}{n},
	\label{eq:compression_complexity}
\end{equation}
where $S_{\mathrm{comp}}$ is the compressed file size in bytes and $n$ is the
number of symbols in the input sequence. The factor~8 converts bytes to bits,
since $1~\text{byte} = 8~\text{bits}$. This expression yields a value in bits per
symbol, providing a dimensionless estimate of algorithmic complexity that is
directly comparable to the entropy rate $h_{\mu}$. Compression libraries such as
\texttt{lzma}, \texttt{bz2}, and \texttt{gzip} report sizes in bytes, making this
normalization necessary for consistency with information-theoretic measures.
Lower values of $C_{\mathrm{comp}}$ indicate greater compressibility and
structural regularity, while higher values reflect increased algorithmic novelty
and unpredictability.

Throughout the remainder of the paper, any reference to reconstruction order or
history length refers to the Markov order $K$. When an optimal order is selected
via model selection criteria, it is denoted by $K^{\ast}$ and used consistently
in all subsequent reconstruction steps unless stated otherwise.

\begin{table}[h]
	\centering
	\caption{Notation used throughout the manuscript.}
	\label{tab:notation}
	\begin{tabular}{ll}
		\toprule
		Symbol & Meaning \\
		\midrule
		$t$ & Discrete simulation time index \\
		$s_t$ & Symbol observed at time $t$ \\
		$\mathcal{A}$ & Symbol alphabet \\
		$L$ & Markov order (history length) \\
		$L^{\ast}$ & Selected Markov order (via BIC) \\
		$L_{\max}$ & Maximum Markov order considered \\
		$s_{t-L:t-1}$ & Length-$L$ symbol history ending at time $t-1$ \\
		$S$ & Causal-state random variable \\
		$\pi_s$ & Stationary probability of causal state $s$ \\
		$P(\cdot \mid s)$ & Emission distribution of state $s$ \\
		$\|\cdot\|_1$ & $\ell_1$ (L1) norm \\
		\bottomrule
	\end{tabular}
\end{table}

\section{Model and data context}

The analysis presented in this paper is based entirely on outputs generated by
an agent-based simulation model. No empirical or observational social data are
used. The purpose of the study is methodological: to demonstrate and evaluate
diagnostic tools for characterizing the temporal organization of agent-based
model (ABM) outputs under controlled simulation conditions.

\subsection*{Agent-based model overview}

The underlying model represents a stylized caregiving system composed of
\emph{caregiver--elder dyads}. Each dyad consists of an older adult requiring
assistance and an informal caregiver providing support. Agents interact within
a simulated environment according to predefined behavioral rules that govern
care provision, mobility, and access to services. Socioeconomic attributes
(e.g., income or functional limitations) are fixed at initialization, while
behavioral and accessibility-related variables evolve over simulation time.

The model is designed to generate heterogeneous agent-level trajectories rather
than to reproduce or fit empirical observations. Simulation outputs are treated
as realizations of a complex adaptive system, suitable for exploratory analysis
of temporal structure and informational organization.

\subsection*{Simulation outputs and variables}

Each simulation run produces time-stamped records for every caregiver--elder
dyad. A dyad is the basic unit of analysis throughout the paper. For each dyad
$i$ and simulation time step $t$, the model records several scalar variables.
The present study focuses on the following outputs:

\begin{itemize}
	\item \textbf{Caregiving effort} (\texttt{efforts}): a continuous-valued
	indicator of the caregiving load or effort exerted by the caregiver at a
	given time step.
	
	\item \textbf{Walkability index} (\texttt{wkb}): a scalar measure of spatial
	accessibility that summarizes the ease with which the dyad can reach
	essential services within the simulated environment.
	
	\item \textbf{Hours not cared} (\texttt{hrsncared}): a non-negative variable
	representing unmet caregiving demand during a given time interval; this
	variable is zero-inflated by construction.
	
	\item \textbf{Overwhelmed status} (\texttt{overwhelmed}): a binary indicator
	signaling whether the caregiver is classified as overloaded at a given time
	step according to model-specific thresholds.
\end{itemize}

All variables are generated endogenously by the simulation rules and are
observed at a fixed temporal resolution.

\subsection*{Meaning of spatial structure}

References to \emph{spatial} or \emph{spatio-temporal} structure in this paper
refer specifically to the simulated physical environment in which agents move
and interact. Spatial structure arises from the representation of locations,
distances, and service accessibility within the model, and from the movement of
agents across this environment over time. The walkability index captures the
dynamic interaction between agent mobility and the spatial configuration of
services. No abstract or metaphorical notion of space is implied.

\subsection*{Analytical scope}

All analyses treat the recorded time series as simulation outputs to be
diagnosed rather than as empirical signals to be explained or predicted. The
goal is to characterize whether, and to what extent, different model variables
exhibit temporal organization, memory, or redundancy under various symbolic
representations. Any conclusions drawn are conditional on the simulation design
and do not constitute claims about real-world caregiving systems.

\section{Proposed Framework for Analyzing ABM Dynamics}
\label{sec:framework}

\subsection{Global $\varepsilon$-Machine as System-Level Reference}
A single $\varepsilon$-machine is reconstructed from a pooled, symbolized time
series (e.g., \texttt{efforts}) using quantile-based discretization. The optimal
Markov order is selected by the Bayesian Information Criterion (BIC) up to
$L_{\max}$, and histories are clustered using the $\ell_1$ norm to obtain causal
states. The resulting $(h_{\mu}, C_{\mu}, E)$ summarize the system-wide
informational regime. When BIC selects $L=0$, the pooled process behaves
approximately memoryless under the chosen resolution, providing a succinct
contextual baseline for finer-grained analyses. By \emph{pooled process} we mean the concatenation of symbolized time series
across all dyads along the temporal axis, after applying a common
discretization scheme. This pooled sequence is used exclusively to construct a
coarse, system-level reference model; it is not intended to represent agent-level
dynamics, nor does it assume ergodicity or exchangeability across dyads.

\subsection{Per-Dyad, Multi-Variable $\varepsilon$-Machines and Algorithmic Proxies}
Each dyad is treated as an independent stochastic process. For each
variable---\texttt{efforts}, \texttt{wkb}, \texttt{hrsncared} (with a hurdle at
zero), and binary \texttt{overwhelmed}---we reconstruct one $\varepsilon$-machine
per dyad, defining a \emph{complexity signature} and obtaining $\{h_{\mu}^{(i,v)}, C_{\mu}^{(i,v)}, E^{(i,v)}\}$. These measures are complemented by
algorithm-agnostic proxies: (i) normalized LZ78 complexity, and (ii) bits per
symbol from lossless compression (LZMA, BZ2, GZIP). These proxies remain
informative even when $\varepsilon$-machines collapse to trivial structures and
provide clusterable feature sets for comparative and multivariate analysis.

\medskip
\noindent
\textbf{Predictive versus descriptive complexity.}
$\varepsilon$-machines and compression-based measures quantify fundamentally
different properties of symbolic processes. $\varepsilon$-machines characterize
\emph{predictive equivalence classes of histories} and therefore capture causal
memory and forward-looking structure. In contrast, compression-based measures
(LZ78, LZMA, BZ2, GZIP) quantify \emph{description length} and redundancy, which
need not imply predictability or causal organization. A sequence may be highly
compressible yet weakly predictive, or vice versa. Their joint use is therefore
diagnostic rather than redundant: agreement indicates robust structure, while
divergence reveals representation- or scale-dependent effects.

\subsection*{Relation to Alternative Reconstruction Methods}

Several alternative approaches exist for reconstructing predictive state models
from symbolic time series. Notably, the Causal State Splitting Reconstruction
(CSSR) algorithm, variable-length Markov models (VLMMs), and related subtree
merging procedures provide principled means to infer state structure under
assumptions similar to those underlying $\varepsilon$-machines.

These methods are widely used in computational mechanics and sequence modeling.
However, they typically require larger sample sizes to stabilize state
splitting, rely on multiple tuning parameters (e.g., significance thresholds or
tree-pruning criteria), and produce reconstructions whose intermediate steps are
less transparent when applied to short or heterogeneous sequences.

In the present study, the analytical objective is not to optimize predictive
performance or to recover a unique generative model, but to \emph{diagnose} the
presence or absence of temporal organization across many short agent-level
trajectories under controlled symbolic representations. For this reason, we
adopt a direct history-clustering formulation of $\varepsilon$-machine
reconstruction with explicit Markov-order selection and $\ell_1$-norm-based
equivalence criteria.

This choice prioritizes interpretability, robustness under limited sample
lengths, and comparability across dyads and variables. Alternative reconstruction
algorithms are therefore excluded by design rather than by omission, and their
use would not alter the qualitative role played here by $\varepsilon$-machines
as diagnostic summaries of predictive structure.

\subsection{Comparative Overview of LZMA, BZ2, and GZIP Compression Algorithms}

The compression algorithms LZMA, BZ2, and GZIP all originate from the
\emph{Lempel--Ziv family} of lossless compression schemes, yet they differ
substantially in their modeling depth, computational efficiency, and ability to
detect long-range dependencies. When applied as proxies for algorithmic
complexity, these differences translate into distinct sensitivities to structural
regularity in symbolic sequences produced by agent-based models.

\medskip
LZMA (Lempel--Ziv--Markov chain algorithm) employs range coding and Markov-chain
modeling to capture both local and distant symbol dependencies. It generally
achieves the highest compression ratios among the three methods, albeit at a
significant computational cost. This makes it particularly appropriate for
sequences with complex or hierarchical patterns, where predictive dependencies
extend over long horizons. Its precision, however, comes at the expense of speed
and memory consumption, which may be nontrivial for large-scale simulations.

\medskip
BZ2, based on the Burrows--Wheeler transform, operates through block sorting
followed by move-to-front and Huffman encoding. It provides an effective
compromise between compression strength and computational cost, being capable of
detecting medium-range regularities without the substantial resource demands of
LZMA. Its performance tends to be stable across moderately redundant or
quasi-periodic data, although it becomes less efficient for short or highly
stochastic sequences.

\medskip
GZIP, which implements the DEFLATE algorithm combining LZ77 windowed matching
with Huffman coding, emphasizes processing speed and portability. It achieves
lower compression ratios but operates efficiently even on small data fragments.
Because of its shallow dictionary window, GZIP predominantly captures short-term
correlations, making it a fast but coarse estimator of descriptive regularity.

\medskip
When interpreted within a complexity-theoretic framework, these algorithms offer
complementary perspectives. The normalized measure is defined as:
\begin{equation}
	C_{\mathrm{comp}} = \frac{8\, S_{\mathrm{comp}}}{n},
	\label{eq:compression_proxy}
\end{equation}
where $S_{\mathrm{comp}}$ is the compressed file size in bytes and $n$ is the
sequence length. This expression quantifies the bits per symbol required to
encode the data. LZMA thus approximates long-range structural predictability,
BZ2 captures intermediate regularities, and GZIP provides an estimate dominated
by local redundancy. Using multiple codecs within the same analytical framework
yields a \emph{multi-scale assessment} of algorithmic regularity.

\subsection{Hierarchical Scopes and Visualization}
Metrics are reported at the dyad (micro) and global (macro) levels. The framework
also supports meso-level aggregation (by stage, disability, or other strata),
enabling future cross-stratum comparisons that summarize heterogeneity across
subgroups. For dissemination, we include (i) distributions and heatmaps of
$\{h_{\mu}, C_{\mu}, E, \mathrm{LZ78}, \mathrm{bps}\}$ by variable and dyad, and
(ii) a concise global $\varepsilon$-machine diagram as a macro-level reference.
This multi-scale representation preserves heterogeneity while providing an
interpretable, system-wide informational baseline.

Throughout this section and the remainder of the paper, references to history
length or reconstruction order refer to the Markov order $L$, while the notation $\|\cdot\|_1$ denotes the $\ell_1$ norm. For a given Markov order $L$, we denote by $s_{t-L:t-1}$ the length-$L$ history of symbols immediately preceding time $t$. When an optimal
order is selected via BIC, it is denoted by $L^{\ast}$ and used consistently in
subsequent analyses unless stated otherwise.

\section{Case Study Outline: Caregiver--Elder Dyads}
Each record in the dataset corresponds to a distinct access event by a caregiver--elder dyad at a given simulation tick. The composite key (\texttt{id\_caregiver}, tick, day, hour) uniquely identifies observations. For each dyad $i$, we consider the multivariate series

\begin{equation}
	X^{(i)}_t = \{\texttt{efforts}, \texttt{wkb}, \texttt{hrsncared}, \texttt{overwhelmed}, \ldots\}_{i,t}.
	\label{eq:multivariate_series}
\end{equation}

Sequences are discretized via quantiles (with a hurdle at zero for \texttt{hrsncared}) and analyzed per variable. Contextual covariates (e.g., age, mobility, disability) are reserved for stratification. The dual analysis yields (a) dyad-level $\varepsilon$-machine measures and (b) algorithmic complexity proxies, enabling stratified comparisons (mobility, occupation, stage) and cohort-level clustering.


The initial $\varepsilon$-machine reconstructions performed with simple \emph{quantile-based} symbolization ($k=5$) illustrate how the discretization of continuous data can decisively shape the inferred informational structure. 
Under this scheme, all three variables---\emph{efforts}, \emph{walkability} (\emph{wkb}), and \emph{hours not cared} (\emph{hrsncared})---collapsed into trivial $\varepsilon$-machines, each with a single causal state ($L=0$, $C_{\mu}=0$) and no measurable predictive information ($E_{\mathrm{proxy}}=0$). 
Although entropy rates remained nonzero ($h_{\mu} \approx 1.37$--$2.32$~bits per symbol), these values merely reflected random variation rather than structured dependence.

\medskip
At first glance, one might conclude that the simulated system is entirely memoryless. 
However, such an interpretation would conflate a \emph{true absence of dynamics} with the \emph{failure of the symbolization} to capture them. 
The discrepancy between the $\varepsilon$-machine outcomes ($E_{\mathrm{proxy}}=0$) and the block-entropy estimates ($E_{\mathrm{block}} \approx 2.3$--$2.9$~bits for \emph{wkb} and \emph{hrsncared}) already signals a deeper issue: 
the symbolization scheme is generating artificial long-range correlations while erasing short-range organization. 
This occurs because quantile binning partitions values according to their rank in the distribution, ignoring the temporal geometry of transitions between them. 
The resulting sequences treat adjacent values that belong to different bins as abrupt categorical changes, even when the underlying process evolves smoothly. 
In such conditions, entropy grows, but the algorithm cannot infer stable predictive states, producing an apparent paradox of \emph{high randomness but no structure}.

\medskip
From a methodological standpoint, this demonstrates why symbolization cannot be regarded as a neutral preprocessing step. 
The quantile scheme was adequate for \emph{efforts}, whose variability is driven primarily by cross-sectional attributes (income and elder ability) rather than by temporal feedback. 
In that case, the absence of causal states corresponds to genuine statistical independence. 
Yet for \emph{walkability} and \emph{hours not cared}, the same discretization distorted the data-generating process: 
it produced $\varepsilon$-machines that appeared memoryless despite substantive reasons to expect autocorrelation---spatial continuity in the former and dependency on caregiving load in the latter. 
More precisely, symbolization defines the observable coarse-graining of the
underlying process. All information-theoretic quantities reported here are
therefore conditional on the chosen representation, in the same sense that
entropy, mutual information, and predictability are conditioned on a
coarse-graining in statistical physics and information theory. Different
symbolizations expose different aspects of the same underlying dynamics:
coarse partitions may suppress short-range dependencies, while finer partitions
may fragment trajectories and inflate entropy. The $\varepsilon$-machine reconstruction
operates on the induced symbolic process and is agnostic to the physical scale
at which observations are discretized.

The broader implication is that symbolization defines the effective topology of information available for analysis. 
A discretization that partitions data by frequency but disregards temporal or contextual continuity will obscure any process that encodes information in the \emph{ordering} of observations rather than in their \emph{amplitude}. 
This is particularly relevant for agent-based simulations, where variables differ not only in magnitude but also in the nature of their dependence on agents’ interactions and the environment. 
Spatially continuous quantities, such as \emph{walkability}, require symbolizations that preserve directional information (e.g., ordinal patterns), whereas zero-inflated behavioral measures, such as \emph{hours not cared}, are better served by hybrid schemes that explicitly distinguish between structural zeros and positive variability (e.g., Hurdle--GMM encodings).

\medskip
In information-theoretic terms, symbolization defines the mapping from the continuous dynamics of the model to the discrete alphabet used for estimating entropy and reconstructing causal states. 
Coarse symbolization merges distinct regimes into single symbols, suppressing predictive information and underestimating complexity; excessively fine symbolization fragments the process, producing spurious states and inflated entropy. 
The objective, therefore, is to identify a symbolization that stabilizes the predictive information ($E_{\mathrm{proxy}}$) while minimizing unnecessary state proliferation. 
In the present analysis, the quantile-based scheme failed to meet this criterion, yielding statistically inconsistent estimates across methods and masking potential dependencies that emerge only under more semantically appropriate encodings.

\section{System-Level $\varepsilon$-machine Reconstruction and Implementation}
\label{sec:implementation}

This section applies the methodological steps summarized in
Section~\ref{sec:framework}. Detailed algorithmic definitions, including
$\varepsilon$-machine construction, Bayesian Information Criterion (BIC)
derivations, finite-block entropy estimators, and the full symbolization and
discretization pipeline, are provided in Appendix~\ref{app:algorithms}.

We report (i) a pooled, system-level reconstruction as a macro reference model
and (ii) dyad-level reconstructions as micro-level \emph{complexity signatures}.
Implementation choices for symbolization, Markov-order selection, clustering
tolerances, and entropy estimation follow the pipeline definitions in
Appendix~\ref{app:algorithms}. Throughout this section, the symbol $L$ denotes
the Markov order, and $\|\cdot\|_1$ denotes the $\ell_1$ norm.

\subsection{System-Level (Pooled) Reconstruction as a Macro Reference}
Let $\mathbf{s}^{\mathrm{pool}} = (s_1,\dots,s_n)$ denote a pooled symbolic
sequence obtained by aggregating the selected variable across dyads and time
after applying a fixed symbolization scheme (Appendix~\ref{app:algorithms}). We
reconstruct a single $\varepsilon$-machine from $\mathbf{s}^{\mathrm{pool}}$ and
report the corresponding $(h_{\mu}, C_{\mu}, E_{\mathrm{proxy}})$ as a compact
summary of the system-wide informational regime under that observational
resolution.

The Markov order $L^{\ast}$ is selected by minimizing $\mathrm{BIC}(L)$ over
$L \in \{0,\dots,L_{\max}\}$, and causal states are obtained by clustering
empirical predictive distributions using the $\ell_1$ norm (see
Appendix~\ref{app:algorithms}). When $L^{\ast}=0$, the pooled process behaves
approximately memoryless under the chosen symbolization. We interpret this case
as a \emph{macro baseline}: it indicates that, at the pooled level and given the
current discretization, predictive structure is not detectable beyond
zero-order dependence, motivating the micro-level analysis below.

\subsection{Dyad-Level Reconstruction and Multi-Variable Feature Extraction}
For each dyad $i$ and each variable $v \in \{\texttt{efforts},\texttt{wkb},
\texttt{hrsncared},\texttt{overwhelmed}\}$, we construct a symbolic sequence
$\mathbf{s}^{(i,v)}$ using the same family of discretization rules
(Appendix~\ref{app:algorithms}) and reconstruct an $\varepsilon$-machine for
$\mathbf{s}^{(i,v)}$. From each reconstructed model we extract the entropy rate,
statistical complexity, and predictive information proxy, yielding the dyad- and
variable-specific signature
\[
\big(h_{\mu}^{(i,v)},\, C_{\mu}^{(i,v)},\, E_{\mathrm{proxy}}^{(i,v)}\big).
\]
These quantities enable distributional analysis across dyads, stratified
comparisons by exogenous attributes (e.g., stage, disability), and unsupervised
clustering of behavioral regimes.

\subsection{Algorithmic Complexity Proxies from Lossless Compression}
Because $\varepsilon$-machine reconstructions may collapse to trivial structures
for short or weakly dependent sequences, we complement the causal-state
quantities with algorithm-agnostic proxies for descriptive regularity. For each
symbolic sequence $\mathbf{s}^{(i,v)}$, we compute: (i) normalized LZ78
complexity; and (ii) bits-per-symbol estimates from multiple lossless codecs
(LZMA, BZ2, GZIP), defined as
\begin{equation}
	C_{\mathrm{comp}} = \frac{8\,S_{\mathrm{comp}}}{n},
	\label{eq:comp_proxy_impl}
\end{equation}
where $S_{\mathrm{comp}}$ is the compressed size in bytes and $n$ is the
sequence length. Using multiple codecs provides a practical multi-scale view of
algorithmic regularity, since LZMA tends to capture longer-range structure, BZ2
captures medium-range regularities, and GZIP is dominated by short-range
redundancy (Appendix~\ref{app:algorithms}).

\subsection{Outputs, Reporting Scope, and Reproducible Artifacts}
All reported metrics are produced at two hierarchical scopes: (i) the pooled
system level and (ii) the dyad level, with optional aggregation to meso-level
strata. For dissemination and auditability, we provide: (a) distributions and
heatmaps of $\{h_{\mu}, C_{\mu}, E_{\mathrm{proxy}}, \mathrm{LZ78},
C_{\mathrm{comp}}\}$ by variable and dyad; (b) a concise diagram of the pooled
$\varepsilon$-machine as a macro reference; and (c) tabular summaries of
$L^{\ast}$ and model size ($|S|$) for each reconstruction. All algorithmic
definitions and implementation details needed to reproduce these outputs are
given in Appendix~\ref{app:algorithms}.

\section{Case Study Results}
\label{sec:casestudy}
\subsection{Global $\varepsilon$-Machine}
\label{sec:structural-signatures}
The symbolic reconstruction of time series derived from the \emph{efforts}, \emph{hours not cared}, and \emph{walkability} variables provides complementary insights into the internal organization of the simulated caregiving system. 
Although these three indicators are jointly produced by the same set of agents, they differ markedly in the degree of temporal structure and informational complexity they encode. 
By combining $\varepsilon$-machine reconstruction with block-entropy estimation, it becomes possible to distinguish between variables that are primarily \emph{distributional}---reflecting cross-sectional variation among agents---and those that exhibit \emph{temporal coherence}, that is, recurrent configurations of states and feedback across simulation steps.

\medskip
For the caregiver effort variable, the $\varepsilon$-machine collapses into a single causal state with no measurable statistical complexity ($C_{\mu}=0$) and a moderate entropy rate ($h_{\mu} \approx 1.37$). 
This outcome is consistent with a process that fluctuates randomly around fixed socioeconomic constraints. 
In the model, caregiving effort is largely determined by the dyad’s income and the elder’s capacity to perform activities of daily living (ADL), both of which are static attributes rather than evolving states. 
Consequently, each observed episode of effort represents an independent draw from a stationary distribution rather than a temporally conditioned decision sequence. 
The absence of predictive information ($E_{\mathrm{proxy}}=0$) confirms that the process is memoryless: it reacts to the current configuration of resources and needs without encoding past experience.

\medskip
A similar pattern characterizes the hours of care not provided. 
Despite a slightly higher entropy rate ($h_{\mu} \approx 1.9$--$2.3$), the process again reduces to a single causal state, indicating that variability is stochastic rather than structured. 
The introduction of a Hurdle--Gaussian Mixture symbolization slightly improved the descriptive fit, but the underlying informational structure remains negligible. 
In substantive terms, this measure captures short-term mismatches between caregiving capacity and demand---episodes that arise independently across simulation ticks rather than as part of a cumulative trajectory. 
The low $C_{\mu}$ and minimal $E_{\mathrm{proxy}}$ thus reflect a process governed by cross-sectional heterogeneity rather than by endogenous feedback or adaptation.

\medskip
By contrast, the walkability variable exhibits a markedly different informational signature. 
Under ordinal symbolic encoding (embedding dimension $m=3$--$4$), the reconstructed $\varepsilon$-machines reveal between 12 and 20 causal states, with complexity values in the range $C_{\mu} \approx 3.3$--$3.6$ and positive predictive information ($E_{\mathrm{proxy}} = 0.3$--$1.2$). 
The entropy rate remains high but stable ($h_{\mu} \approx 2.1$--$3.3$), indicating the coexistence of stochastic and deterministic elements. 
This configuration is typical of processes in which spatial continuity and behavioral feedback jointly generate structured but non-periodic dynamics. 
In this model, \emph{walkability} synthesizes the influence of infrastructure, terrain, and service distribution with the movement routines of the dyads. 
As agents repeatedly interact with a constrained environment, localized accessibility patterns emerge, generating recurrent---though not fully deterministic---mobility regimes. 
The resulting $\varepsilon$-machine suggests a bounded but genuine temporal memory, in which present states retain information about prior accessibility conditions.

\medskip
Taken together, these results delineate a clear distinction between static socioeconomic constraints and dynamic spatial interactions within the simulated caregiving environment. 
The first two variables (\emph{efforts} and \emph{hours not cared}) behave as conditionally independent draws shaped by exogenous attributes; 
the third (\emph{walkability}) reflects an endogenous spatio-behavioral subsystem capable of storing and transmitting information across time. 
In informational terms, the caregiving process appears \emph{structurally flat} in its socioeconomic dimension but \emph{dynamically rich} in its spatial and mobility component. 
This finding supports the interpretation that infrastructural and environmental factors---rather than individual or household resources---constitute the principal carriers of systemic complexity and feedback in the simulated model of care accessibility.

\medskip
From a methodological standpoint, the contrast among these variables illustrates the diagnostic value of $\varepsilon$-machine reconstruction for agent-based simulations. 
When applied to outputs derived from mixed socioeconomic and spatial processes, computational-mechanics metrics provide a compact representation of where and how information is stored in the system. 
In this case, the emergence of multiple causal states in \emph{walkability} but not in \emph{effort} or \emph{care deficit} time series indicates that spatial interactions operate as the primary generator of temporal organization. 
Future extensions could explicitly couple these subsystems---for example, allowing caregiver fatigue or service accessibility to evolve as a function of prior overload or mobility episodes---to test whether feedback between socioeconomic and spatial dimensions increases the system’s overall statistical complexity and predictive capacity.

\begin{figure}[H]
	\centering
	\includegraphics[width=0.32\textwidth]{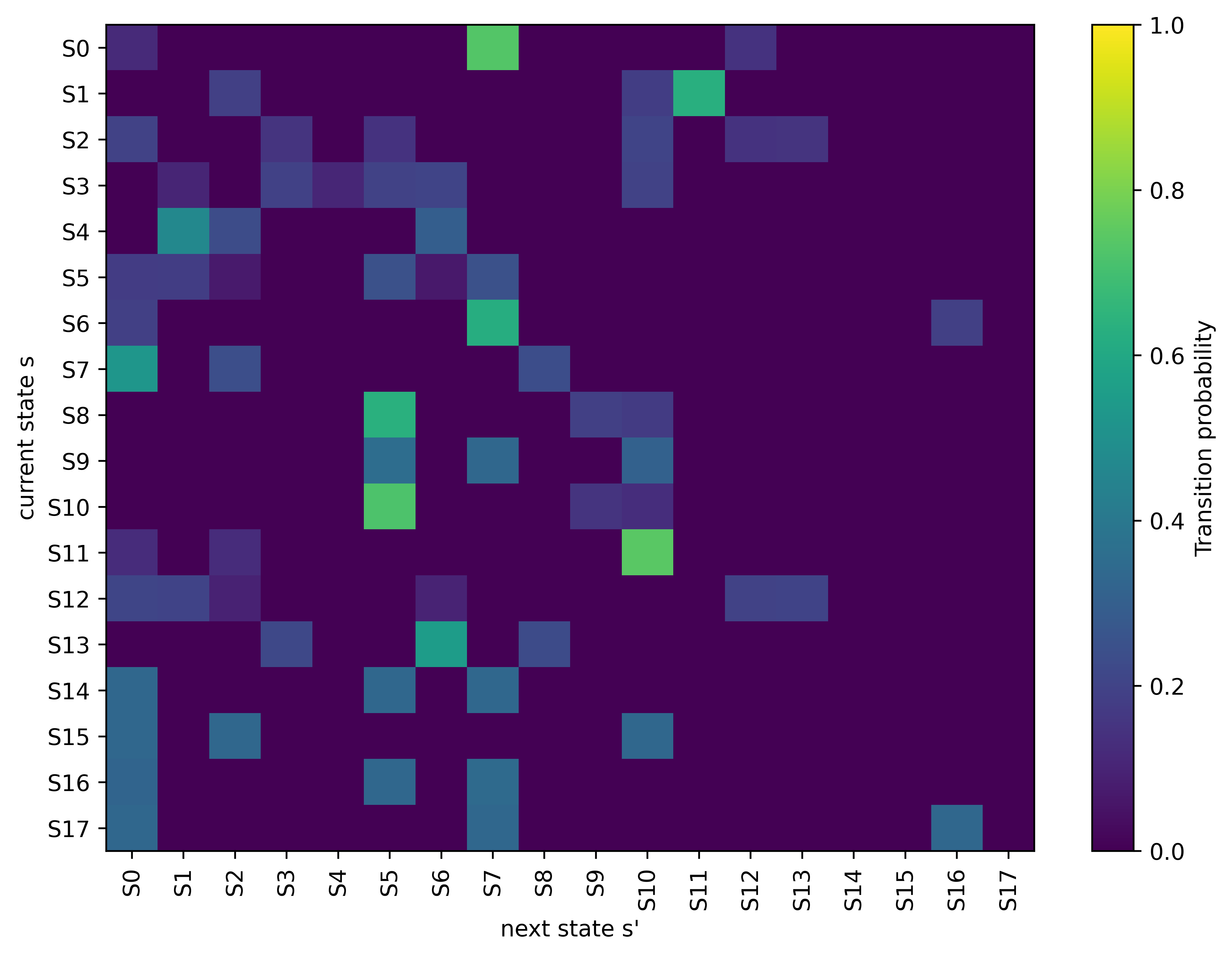}
	\includegraphics[width=0.32\textwidth]{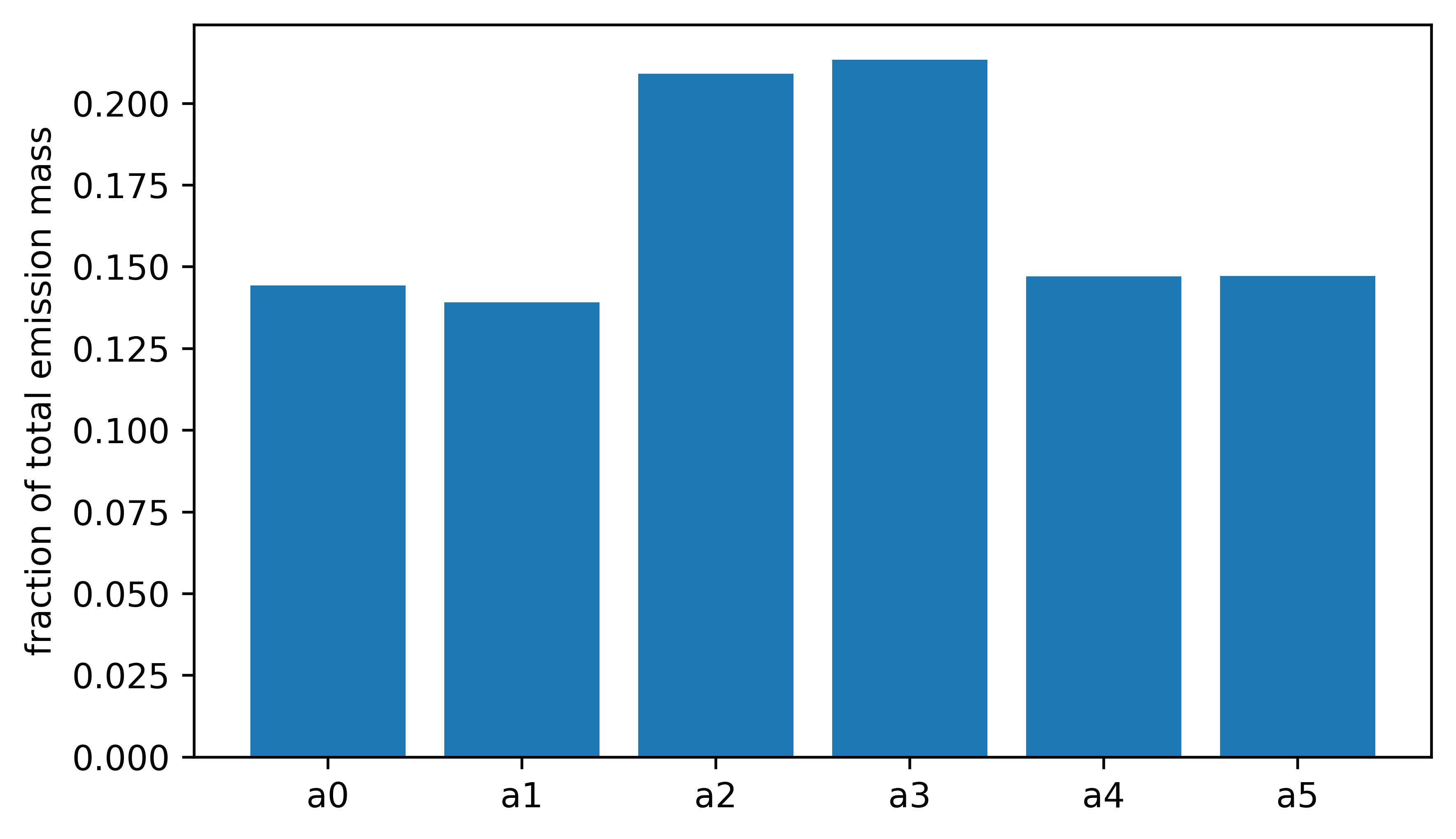}
	\includegraphics[width=0.32\textwidth]{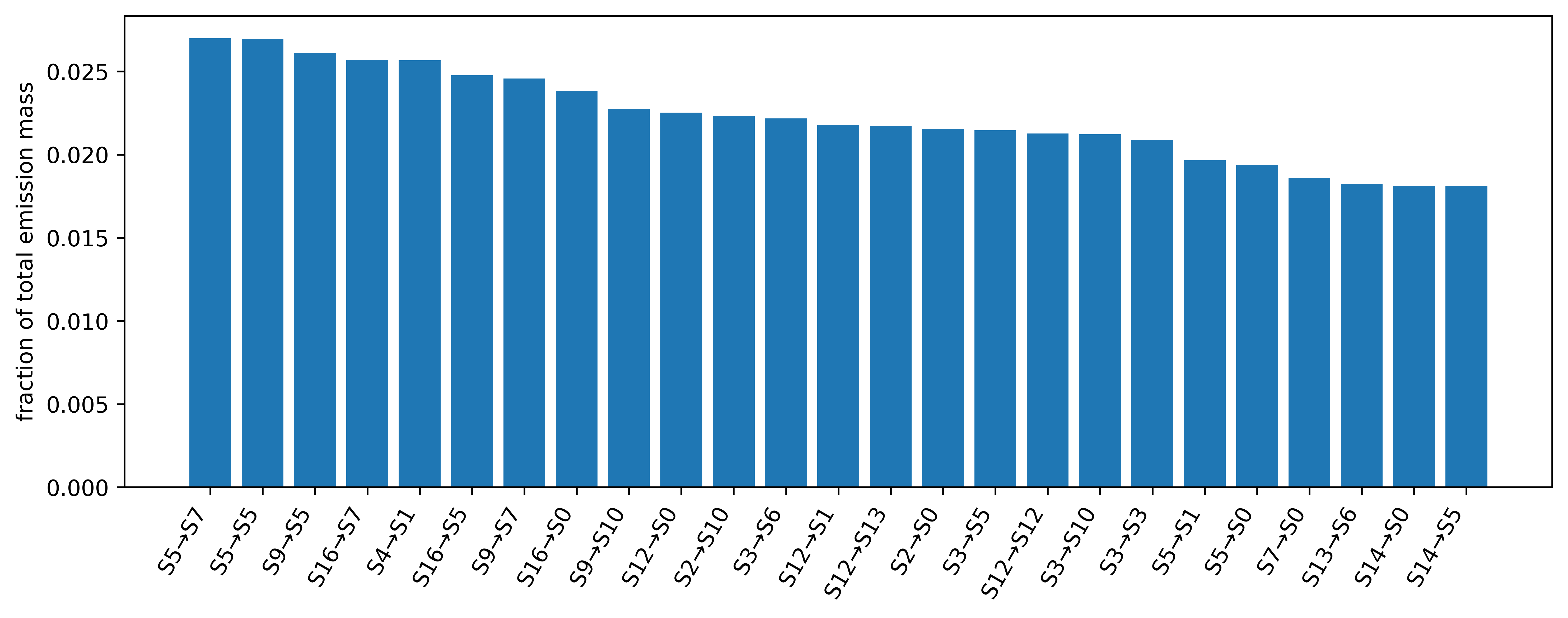}
	\caption{Structural signatures. Left: 	State--transition probability matrix.
		Heatmap of transition probabilities between causal states (rows = current $s$, columns = next $s'$). 
		Bright diagonal elements indicate persistence within the same state, whereas off-diagonal clusters denote systematic switching among subsets of states. 
		The resulting pattern highlights modular organization and localized predictability within the global $\varepsilon$-machine. Emission and transition mass distributions. 
		Middle: Fraction of total emission probability associated with each ordinal symbol ($a_0$--$a_5$). 
		Middle-range motifs ($a_2$, $a_3$) dominate, suggesting gradual rather than abrupt changes in relative walkability. 
		Right: Most probable state-to-state transitions (top 25 by mass). 
		The concentration of probability mass among a few edges (e.g., $S_5 \!\to\! S_7$, $S_5 \!\to\! S_5$) indicates limited recurrent micro-patterns and short-range stability.}
	\label{fig:structural-signatures}
\end{figure}

\begin{figure}[H]
	\centering
	\includegraphics[width=0.475\textwidth]{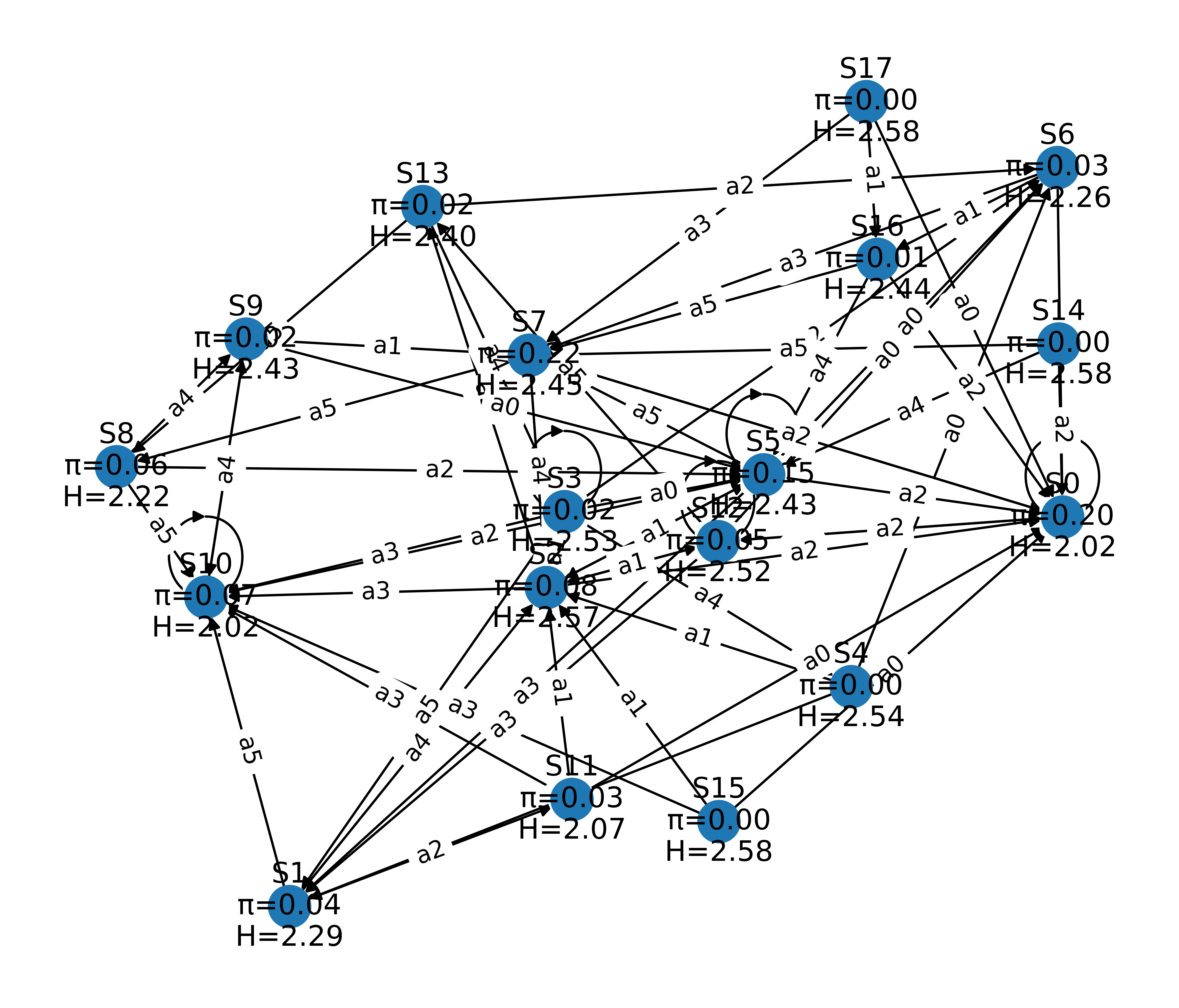}
	\caption{Global $\varepsilon$-machine for walkability (ordinal $m=3$). 
		Directed graph representation of the reconstructed $\varepsilon$-machine. 
		Nodes correspond to causal states ($S_0$--$S_{17}$) annotated with their stationary probability ($\pi$) and local emission entropy ($H$). 
		Edges denote predictive transitions labeled by emitted ordinal symbols ($a_0$--$a_5$). 
		The layout reveals a moderately connected topology with clusters of recurrent states and several self-loops, indicating locally stable yet globally fragmented dynamics.}
	\label{fig:global-epsm}
\end{figure}

The directed graph representation of the reconstructed $\varepsilon$-machine provides a visual summary of the system’s internal information architecture. 
Each node represents a \emph{causal state}, labeled by its stationary probability ($\pi$) and local emission entropy ($H$), while directed edges encode transitions labeled by emitted ordinal symbols ($a_0$--$a_5$). 
The resulting topology captures how past configurations of the process are used to generate its future evolution. 
In the global $\varepsilon$-machine for \emph{walkability} ($m=3$), the network exhibits a moderately connected structure with a limited number of high-probability nodes and recurrent loops. 
This pattern indicates that the system’s predictive organization is \emph{localized}---information tends to circulate within small state clusters rather than propagate through long transition chains. 
Such a structure implies short-range predictability and context-bound adaptation rather than persistent global memory.

\medskip
Complementary visualizations of the same $\varepsilon$-machine help to uncover additional layers of structure. 
The state--transition heatmap acts as a \emph{structural signature} of the reconstructed process: each cell encodes the empirical probability of transitioning from state $i$ to state $j$ within either the global or per-dyad symbolic sequence.

\begin{itemize}
	\item Diagonal dominance denotes \emph{persistence}: agents or dyads tend to remain in similar ordinal configurations across successive time steps, suggesting that local walkability (or caregiving effort) states are self-reinforcing.
	\item Off-diagonal clusters or bands reveal \emph{systematic switching}, that is, recurring transitions between specific subsets of states, often corresponding to alternating behavioral or spatial regimes (e.g., shifts between high- and low-accessibility contexts or cyclical caregiving routines).
\end{itemize}

This two-dimensional compression of the full $\varepsilon$-machine makes nonlinear dependencies and \emph{predictive symmetry} visually salient, facilitating cross-comparison of simulations under different contextual or policy conditions.

\medskip
The symbol histogram summarizes the marginal distribution of ordinal patterns---here, the six motifs $a_0$--$a_5$ generated for $m=3$. 
It effectively captures the \emph{compositional balance} of local dynamics.

\begin{itemize}
	\item A \emph{flat} histogram signals near-equiprobable symbol usage, implying unstructured or noise-like fluctuations.
	\item Conversely, \emph{skewed} or \emph{multimodal} distributions indicate dominance of particular directional motifs (e.g., monotonic increases or decreases in walkability), exposing \emph{asymmetric dynamical tendencies} in the system.
\end{itemize}

This representation also provides an empirical check on the entropy-rate estimates ($h_{\mu}$): strongly uneven symbol frequencies should coincide with lower entropy and hence greater short-range predictability.

\medskip
Finally, the state-pair histogram aggregates transitions between causal states rather than symbols. 
It functions as a \emph{coarse projection} of the $\varepsilon$-machine graph, emphasizing which causal transitions dominate the overall process without displaying the entire network.

\begin{itemize}
	\item High-probability self-loops identify stable, quasi-stationary behavioral regimes.
	\item Concentrated transitions between a few state pairs delineate a \emph{low-dimensional predictive core}, where the system’s dynamics can be summarized by a small number of recurrent pathways.
	\item Dispersed transition weights, by contrast, indicate \emph{context-dependent or weakly structured memory}, a finding consistent with the relatively low $E_{\mathrm{proxy}}$ values estimated for most dyads.
\end{itemize}

\subsubsection{Interpretation}

The reconstructed $\varepsilon$-machine for \emph{walkability}, obtained under an ordinal encoding with embedding dimension $m=3$, reveals a moderately complex yet sparsely connected predictive structure. The causal-state network contains eighteen recurrent states, most of which exhibit small stationary probabilities and moderate local emission entropy ($\approx 2$--$2.6$~bits). The configuration points to a process that is locally regular but globally weakly coupled: individual dyads experience recurrent short-term mobility patterns, whereas transitions among those regimes remain infrequent.

The symbol-level distributions further clarify the system’s organization. The predominance of middle-rank ordinal motifs ($a_2$, $a_3$) indicates that, for most dyads, changes in walkability occur gradually rather than through abrupt jumps. 
Extreme orderings ($a_0$, $a_5$) are comparatively rare, suggesting that accessibility fluctuates within a limited dynamic range. Similarly, the transition-mass histogram demonstrates that a small subset of causal-state transitions concentrates most of the emission probability, implying that the overall process revisits a few micro-configurations repeatedly. This pattern is consistent with short-range stationarity and the predominance of local feedback mechanisms within each dyad’s spatial context.

The transition heatmap provides an additional structural perspective. The visible diagonal dominance corresponds to persistence, meaning that once the process occupies a particular causal state, it tends to remain there for multiple steps. At the same time, a few off-diagonal bands reveal recurrent oscillations between paired states---an indication of cyclical local adjustments rather than long-term directional trends. Collectively, these features suggest that the walkability dynamics, as captured at the dyad time scale, are quasi-stationary: they display internal memory and organization but only within limited temporal and spatial neighborhoods.

From a broader perspective, this global $\varepsilon$-machine constitutes a structural signature of the underlying simulated process. The moderate number of states, the partial diagonal dominance of the transition matrix, and the concentration of symbolic mass in a few motifs together depict a system where adaptation and inertia coexist. Walkability patterns are thus better described as \emph{locally persistent and self-referential} rather than globally predictive. Detectable memory likely emerges only after conditioning on slower or aggregated temporal structures---such as daily or weekly rhythms---or through multivariate coupling with mobility and caregiving variables.

\subsection{Per-Dyad / Per-Variable $\varepsilon$-Machine Reconstruction with Kolmogorov-Style Complexity Proxies}

While the global model (pooled series, single $\varepsilon$-machine) summarizes the system-level informational architecture through $(h_{\mu},\, C_{\mu},\, E)$ and provides a useful baseline—for instance, $L=0$ implies that the pooled process is approximately memoryless under the chosen symbolic resolution—the per-dyad models recover idiosyncratic predictive structures and variability across agents and variables, with algorithmic proxies offering an orthogonal view of compressibility. 
Together, these representations separate semantic organization (via the $\varepsilon$-machine: predictive causation and memory) from syntactic simplicity (via compression: description length), yielding a unified lens on emergence, adaptation, and heterogeneity in agent-based model dynamics.

\subsubsection{Compression-Based Diagnostics of Structural Complexity}

Compression metrics provide an interpretable and low-dimensional summary of emergent structure in the simulation. 
The joint examination of statistical and algorithmic compressibility enables one to distinguish randomness from organized variability, to classify processes by stability versus intermittency, and to assess redundancy among symbolic representations. 
Such compression-based diagnostics therefore complement the $\varepsilon$-machine analysis by linking symbolic dynamics to computable measures of order, offering a quantitative signature of systemic organization that can be tracked across scenarios or policy interventions. 

\medskip
Specifically, we quantify statistical and algorithmic regularities across simulated variables (\emph{efforts}, \emph{wkb}, \emph{hrsncared}). 
These indices capture different aspects of informational organization: the LZ78 normalized complexity (\texttt{lz78\_norm}) reflects statistical unpredictability based on substring diversity, while the LZMA, BZ2, and GZIP \emph{bits-per-symbol} (\texttt{bps}) values estimate algorithmic compressibility—how efficiently sequences can be encoded under general-purpose compression schemes. 
Together, these measures offer a dual view of structural variability: one grounded in symbolic recurrence and the other in algorithmic redundancy.

\paragraph{Summary of $\varepsilon$-Machine and Kolmogorov Metrics.}

\begin{table}[H]
	\centering
	\caption{Descriptive statistics and informational metrics across configurations: sample size per dyad ($n$), alphabet size ($A$), selected Markov order ($L$), number of causal states $|S|$, entropy rate $h_\mu$, statistical complexity $C_\mu$, excess-entropy proxy $E$, and compression-based measures (LZMA/BZ2/GZIP bits-per-symbol; normalized LZ78).}
	\label{tab:epsm_summary}
	\resizebox{\textwidth}{!}{%
		\resizebox{\textwidth}{!}{%
			
\centering
\begin{tabular}{l l c c c c c c c c c}
\toprule
Variable & Symbolization & $\overline{h_\mu}$ & $\overline{C_\mu}$ & $\overline{E}_{\text{proxy}}$ &
LZ78$_\text{norm}$ & LZMA (bps) & BZ2 (bps) & GZIP (bps) & Mode $L$ & Median $|\mathcal{S}|$ \\
\midrule
efforts & value & 1.15 ± 1.01 & 0.01 ± 0.12 & 0.00 ± 0.00 & 0.90 ± 0.42 & 2.13 ± 1.46 & 1.70 ± 1.23 & 1.50 ± 1.15 & 0 & 1 \\
hrsncared & value & 2.01 ± 0.43 & 0.00 ± 0.00 & 0.00 ± 0.00 & 1.16 ± 0.14 & 3.52 ± 0.39 & 2.67 ± 0.34 & 2.52 ± 0.31 & 0 & 1 \\
overwhelmed & binary & 0.03 ± 0.15 & 0.00 ± 0.00 & 0.00 ± 0.00 & 0.45 ± 0.20 & 0.51 ± 0.19 & 0.30 ± 0.22 & 0.20 ± 0.18 & 6 & 1 \\
wkb & ordinal & 2.34 ± 0.81 & 0.13 ± 0.45 & 0.03 ± 0.09 & 0.67 ± 0.15 & 2.55 ± 0.58 & 1.79 ± 0.41 & 2.03 ± 0.54 & 0 & 1 \\
\bottomrule
    \end{tabular}
		}%
		
	}
\end{table}

Table~\ref{tab:epsm_summary} summarizes the main outcomes across value-binned and ordinal encodings and parameter variants. 
Most dyads exhibit limited short-range predictability unless temporal symbolization explicitly encodes ordinal relations.

\paragraph{Symbolization Schemes and Edges.}

\begin{table}[H]
	\centering
	\caption{Symbolization choices and recovered edges by variable. 
		Value-binned variables (\emph{efforts}, \emph{hrsncared}) use global quantile and hurdle schemes; 
		\emph{wkb} (walkability) adopts ordinal encodings ($m{=}3$ or $m{=}4$).}
	\label{tab:symbol_edges}
	\resizebox{0.9\textwidth}{!}{%
		
\centering
\begin{tabular}{l p{0.52\linewidth} p{0.22\linewidth}}
\toprule
Variable & Edges (bin boundaries or descriptor) & Labels \\
\midrule
efforts & \texttt{[0.0, 0.03, 0.19, 0.8]} & \texttt{['q1', 'q2', 'q3', 'q4', 'q5']} \\
hrsncared & \texttt{[0.0, 1e-12, 0.08, 1.26, 3.13, 10.07]} & \texttt{['zero', 'pos\_q1', 'pos\_q2', 'pos\_q3', 'pos\_q4']} \\
wkb &  & \texttt{['ordinal\_m=4 (alphabet=24)']} \\
wkb &  & \texttt{['ordinal\_m=3 (alphabet=6)']} \\
\bottomrule
    \end{tabular}

	}
\end{table}

Table~\ref{tab:symbol_edges} details the discretization edges. 
The balance between granularity and parsimony governs the trade-off between entropy and over-segmentation.

\paragraph{Ordinal Walkability: Model Selection Across Dyads.}

\begin{table}[H]
	\centering
	\caption{Model selection summary for ordinal walkability across dyads and merge thresholds. 
		Conservative merging with $m{=}4$ yields $L{=}0$ for most dyads; 
		reducing to $m{=}3$ and relaxing merges reveals finite memory ($L{=}1$–$2$) for a subset.}
	\label{tab:wkb_model_selection}
	\resizebox{0.9\textwidth}{!}{%
		
	\centering
	\begin{tabular}{lcccc}
		\toprule
		Statistic & \# Dyads & $\Pr(L>0)$ & $\Pr(\text{practical memory})$ & Mean $L$ $\pm$ SD \\
		\midrule
		wkb (ordinal) & 94 & 10.3\% & 8.2\% & $0.21 \pm 0.90$ \\
		\bottomrule
	\end{tabular}

	}
\end{table}

Table~\ref{tab:wkb_model_selection} shows that most dyads collapse to $L{=}0$ under conservative merging, confirming weak short-range memory. 
With $m{=}3$ and relaxed merging, a subset exhibits finite predictive structure ($L{=}1$–$2$) and moderate excess entropy.

\paragraph{Statistical vs.\ Algorithmic Compressibility.}

The scatterplot of \texttt{lz78\_norm} versus \texttt{lzma\_bps} provides a two-dimensional projection of statistical and algorithmic complexity. 
Points aligned along the main diagonal denote sequences that are simultaneously complex in a statistical sense yet compressible algorithmically—structured but not trivially repetitive (frequent for \emph{walkability} and \emph{efforts}). 
Deviations from the diagonal, particularly \textbf{high \texttt{lzma\_bps} with low \texttt{lz78\_norm}}, indicate algorithmic irregularity: symbol sequences that lack local redundancy yet embed context-dependent transitions—typical of intermittent caregiving or fluctuating accessibility where abrupt changes dominate over persistent cycles.

\begin{figure}[H]
	\centering
	\includegraphics[width=0.48\textwidth]{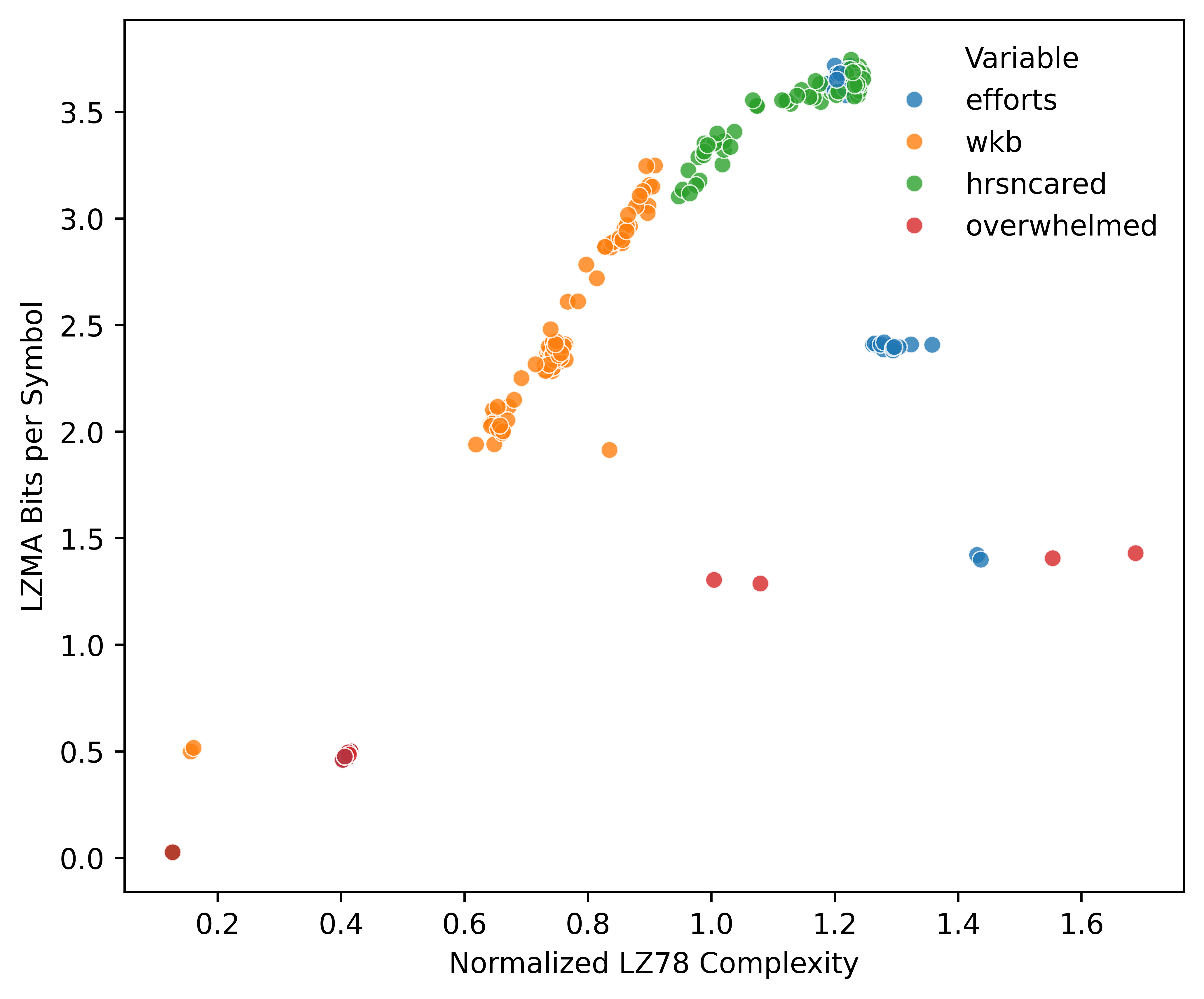}\hfill
	\includegraphics[width=0.48\textwidth]{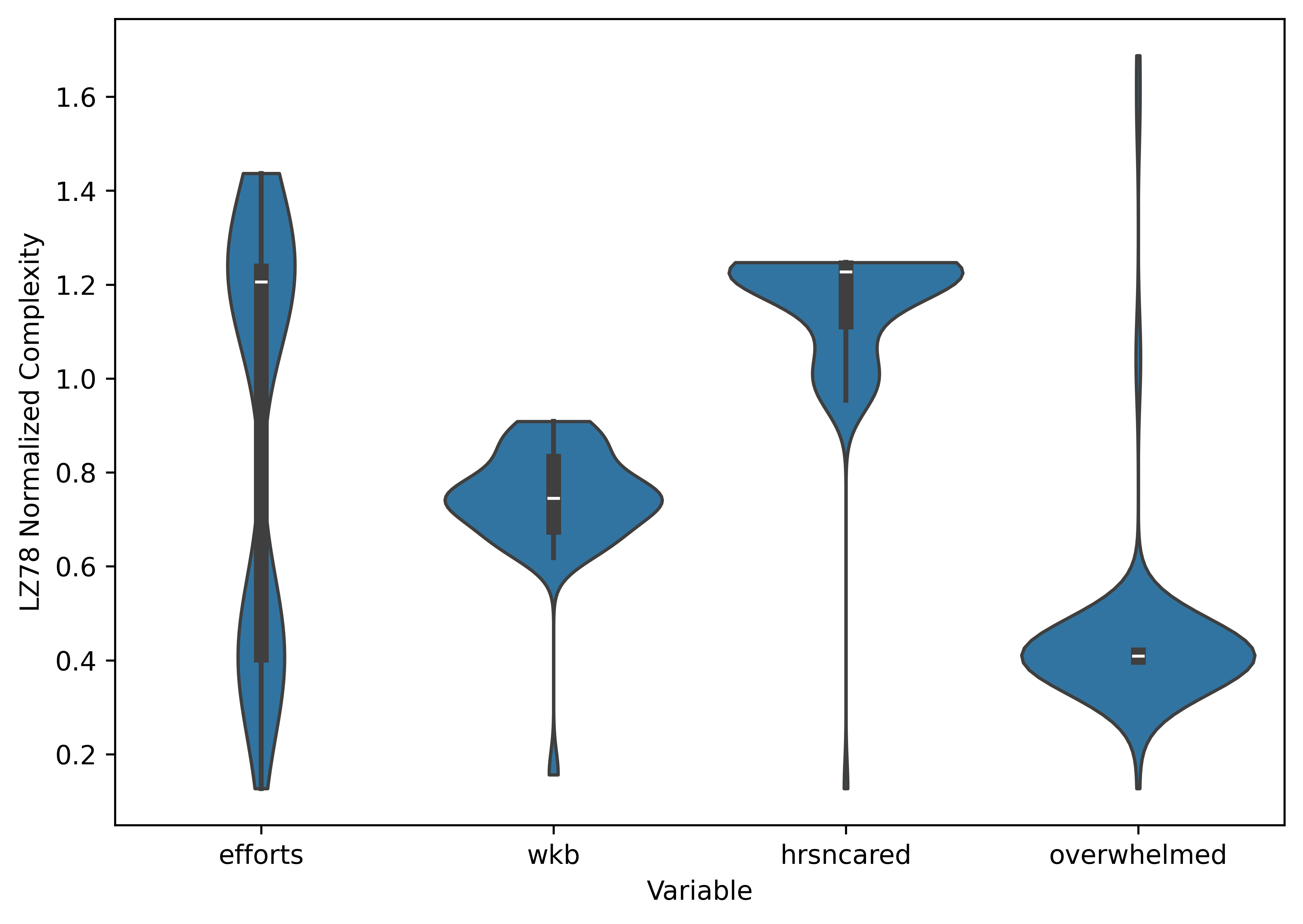}
	\caption{Left: normalized LZ78 complexity vs.\ LZMA bits per symbol (statistical vs.\ algorithmic compressibility). Right: violin plots of normalized LZ78 by variable showing heterogeneity and burstiness.}
	\label{fig:kolmogorov-1}
\end{figure}

\paragraph{Distributional Structure Across Variables.}

Boxplots and violin plots of the compression metrics by variable illustrate distinct informational signatures. 
\emph{Efforts} displays narrow and nearly symmetric distributions, indicating stable, low-variance dynamics consistent with short-range memoryless behavior. 
In contrast, \emph{walkability (wkb)} and \emph{hours not cared} exhibit broader and often asymmetric profiles, suggesting heterogeneous or bursty processes that reflect spatial and caregiving variability within the simulation. 
Overall, the behavioral and spatial dimensions differ in their intrinsic regularity, with \emph{efforts} serving as a quasi-stationary reference and \emph{wkb} capturing higher-order nonlinear variability.

\paragraph{Correlations Among Compression Measures.}

Table~\ref{tab:compression_corr} reports pairwise correlations among the compression-based metrics (\texttt{lzma\_bps}, \texttt{bz2\_bps}, \texttt{gzip\_bps}, and \texttt{lz78\_norm}). 
The results show strong interdependence among dictionary compressors—LZMA, BZ2, and GZIP—(typically $\rho > 0.9$), confirming that these algorithms provide consistent estimates of algorithmic compressibility. 
In contrast, the normalized LZ78 complexity diverges moderately ($\rho \approx 0.7$–$0.8$), indicating that statistical and algorithmic compressibility capture complementary informational dimensions. 
Whereas compression ratios quantify encoding efficiency, normalized LZ78 reflects the rate of novel pattern introduction. 
High statistical complexity combined with moderate algorithmic compressibility typifies structured yet nonreducible dynamics, a hallmark of adaptive, path-dependent agent-based systems.

\begin{table}[H]
	\centering
	\caption{Correlation matrix among compression metrics (LZ78, LZMA, BZ2, GZIP). 
		High inter-algorithm correlations ($>0.9$) indicate redundancy among compressors, 
		whereas moderate divergence from LZ78 suggests complementary sensitivity to symbolic recurrence.}
	\label{tab:compression_corr}
	\resizebox{0.7\textwidth}{!}{%
		
\centering
\begin{tabular}{lrrrr}
\toprule
 & LZMA (bps) & BZ2 (bps) & GZIP (bps) & LZ78$_\text{norm}$ \\
\midrule
LZMA (bps) & 1.00 & 0.98 & 0.99 & 0.85 \\
BZ2 (bps) & 0.98 & 1.00 & 0.98 & 0.90 \\
GZIP (bps) & 0.99 & 0.98 & 1.00 & 0.82 \\
LZ78$_\text{norm}$ & 0.85 & 0.90 & 0.82 & 1.00 \\
\bottomrule
    \end{tabular}

	}
\end{table}

\subsubsection{Interpretation}

The plots jointly characterize the compressibility and structure of dyad-level time series, revealing how the informational content of each behavioral dimension (\emph{efforts}, \emph{walkability index} [\emph{wkb}], \emph{hours not cared}, and \emph{overwhelmed status}) differs when viewed through statistical versus algorithmic lenses. Two main regimes emerge. The first corresponds to a diagonal band in the LZ78–LZMA scatterplots, where statistical and algorithmic compressibility are highly dependent. These sequences are structured yet predictable, indicating organized variability rather than noise. The second regime consists of outliers characterized by high \texttt{lzma\_bps} but low \texttt{lz78\_norm}, reflecting algorithmic irregularity driven by intermittent or context-dependent transitions. Variable-level clustering suggests that behavioral regularity and spatial accessibility interact differently with predictability, depending on whether dynamics are continuous, as in \emph{efforts}, or episodic, as in \emph{hours not cared}.

Normalized LZ78 distributions further differentiate the informational geometry of variables. The \emph{efforts} and \emph{hours not cared} series display higher medians and moderate dispersion, indicative of steady but nuanced dynamics. In contrast, \emph{walkability} shows lower and narrower complexity, constrained by the ordinal structure of the spatial network, while the \emph{overwhelmed} variable exhibits strong asymmetry and occasional multimodality consistent with threshold-type behavior. 

Compression bits-per-symbol derived from LZMA, BZ2, and GZIP confirm these patterns. The wider dispersion observed for \emph{efforts} and \emph{walkability} signals heterogeneous temporal organization, whereas \emph{hours not cared} exhibits narrower interquartile ranges, implying quasi-stationary dynamics once a caregiving rhythm stabilizes. The \emph{overwhelmed} variable, in contrast, presents near-zero median values, reflecting a binary regime with minimal informational entropy.

Correlations among compression metrics reinforce this interpretation. Dictionary-based compressors (LZMA, BZ2, GZIP) show very high mutual dependence ($\rho > 0.97$), indicating they capture overlapping redundancy structures, while their weaker association with normalized LZ78 ($\rho \approx 0.7$–$0.8$) highlights the complementary nature of the two perspectives. Whereas dictionary compressors quantify algorithmic predictability through redundancy elimination, LZ78 emphasizes statistical recurrence and symbolic novelty. Considering both metrics thus provides a robust diagnostic framework for detecting the emergence of complexity and variability in agent-level behavioral trajectories. 

Overall, the results indicate that most dyads exhibit moderate algorithmic structure rather than purely random or deterministic dynamics. Temporal heterogeneity varies systematically across behavioral variables, reflecting diverse cognitive and spatial constraints. Statistical and algorithmic measures, though correlated, are not interchangeable; together they offer a more comprehensive understanding of how predictability and adaptation emerge in the caregiving system.

\section{Discussion, Limitations, and Conclusions}
\label{sec:discussion}

This study applies tools from computational mechanics to the analysis of
agent-based model (ABM) outputs, with the specific goal of diagnosing temporal
organization and predictive structure in simulated social dynamics. While
$\varepsilon$-machine reconstruction has been developed primarily in the context
of symbolic and nonlinear dynamical systems
\cite{CrutchfieldYoung1989,Crutchfield1994,ShaliziCrutchfield2001}, its use in
applied ABM analysis remains limited. The present work demonstrates how these
methods can be operationalized within a data-driven simulation workflow, without
relying on theoretical toy processes or analytically specified sources.

Methodologically, the framework integrates symbolic dynamics, causal-state
reconstruction, and algorithmic compression measures to characterize the
informational organization of ABM-generated time series. Each component targets
a distinct aspect of structure: $\varepsilon$-machines identify predictive
equivalence classes of histories; entropy- and complexity-based quantities
summarize information generation and storage; and compression-based indices
quantify description length and redundancy. Taken together, these diagnostics
provide a reproducible and internally consistent description of temporal
regularity across variables, agents, and observational resolutions.

A central distinction clarified by the analysis is that between
\emph{predictive structure} and \emph{descriptive regularity}. Following
computational mechanics, intrinsic computation concerns how past observations
constrain future behavior through causal states
\cite{Crutchfield1994,ShaliziCrutchfield2001}. Compression-based measures, by
contrast, estimate the compactness of a symbolic description without implying
causal predictability. The joint application of these tools therefore serves a
diagnostic purpose: agreement between them indicates robust structure, whereas
divergence highlights scale-, representation-, or variable-specific effects.

Empirically, the results indicate that several variables produced by the
caregiving ABM---notably \emph{efforts} and \emph{hours not cared}---exhibit
near-memoryless behavior at the dyad time scale under the symbolizations
considered. In contrast, \emph{walkability}, when encoded using ordinal patterns,
retains weak but detectable predictive structure. This suggests that temporal
organization in the model is variable-specific and closely tied to spatial and
interaction-driven processes rather than to purely cross-sectional attributes.
Importantly, the absence of detectable memory at a given resolution does not
imply an absence of structure in the underlying dynamics, but rather reflects
the conditional nature of inference under a chosen coarse-graining.

These findings underscore the role of symbolization as an explicit modeling
assumption. Symbolization defines the observable coarse-graining of the process,
and all reported information-theoretic quantities are conditional on this
representation, in the same sense that entropy and mutual information depend on
the chosen partition of state space in information theory
\cite{Shannon1948,CoverThomas2006}. Ordinal encodings preserve local temporal
geometry, while hybrid schemes such as hurdle--mixture discretizations address
structural zeros. No symbolization is privileged as “correct”; rather, different
encodings expose different aspects of the same simulated dynamics.

\paragraph{Summary of Findings.}
At the system level, pooled reconstructions often select $L^{\ast}=0$ under
coarse symbolizations, yielding a macro-level baseline that is effectively
memoryless. At the dyad level, however, $\varepsilon$-machine reconstructions
reveal heterogeneity across variables and agents, with nontrivial predictive
structure emerging primarily for spatially mediated quantities under ordinal
symbolizations. Compression-based diagnostics corroborate these patterns by
distinguishing between redundancy and novelty across sequences, even when causal
state structure is minimal.

\paragraph{Interpretation.}
The contrast between socioeconomic and spatial variables suggests that temporal
organization in the model arises primarily from interaction with the simulated
environment rather than from static agent attributes. This separation between
predictive memory and descriptive simplicity clarifies how different forms of
structure coexist in ABM outputs and how they can be disentangled using
complementary diagnostics.

\paragraph{Methodological Implications.}
The analysis highlights the importance of reporting multiple, non-redundant
complexity measures. We recommend presenting entropy-rate–normalized quantities,
statistical complexity, predictive-information proxies, and compression-based
bits-per-symbol jointly, to assess robustness across symbolizations and scales.
Reliance on a single metric risks conflating descriptive redundancy with
predictive organization.

\subsection*{Limitations and Scope}

The results presented here are subject to several important limitations. First,
all analyses are simulation-specific. The symbolic sequences analyzed are
generated entirely by an agent-based model and do not constitute empirical
observations. Accordingly, the reported measures characterize the informational
organization of the simulated dynamics under controlled assumptions, rather than
explaining or predicting real-world caregiving behavior.

Second, all information-theoretic quantities are representation-dependent.
Symbolization defines the observable coarse-graining of the underlying process,
and entropy rates, statistical complexity, and predictive information are
conditional on this choice, as is standard in information theory and statistical
physics \cite{shannon1948,CoverThomas2006}. Alternative discretizations may reveal
different aspects of the same underlying dynamics.

Third, the framework is explicitly diagnostic rather than explanatory.
$\varepsilon$-machine reconstructions identify predictive equivalence classes of
histories and quantify temporal organization, but they do not identify causal
mechanisms in an interventionist sense, nor do they constitute generative models
of agent decision-making \cite{Crutchfield1994,ShaliziCrutchfield2001}. The
reported measures should therefore be interpreted as descriptive summaries
suited to comparison, stratification, and robustness analysis.

\paragraph{Future Directions.}
Future work should incorporate uncertainty quantification for entropy,
complexity, and compression-based measures, for example via block bootstrap or
surrogate-based methods. Methodological extensions include multivariate
$\varepsilon$-machines, delayed embeddings to capture slower rhythms, and
symbolizations informed by policy-relevant thresholds. Applying the framework to
nonstationary or intervention-driven ABMs may further clarify how informational
structure responds to structural change in simulated social systems.

\section*{Funding}
This work was supported by the European Union -- Next Generation EU under the National Recovery and Resilience Plan (NRRP) project Age-It (AGE-IT -- PE00000015; CUP: H43C22000840006).

\section*{Data availability}
All simulation code, parameter files, and processed datasets used in this study are available from the corresponding author upon reasonable request. 
The source code for symbolization, $\varepsilon$-machine reconstruction, and complexity proxies (normalized LZ78 and compression-based metrics) will be archived with a versioned hash to ensure reproducibility. 
The dataset schema includes identifiers (\texttt{id\_caregiver}), temporal fields (\texttt{tick}, \texttt{day}, \texttt{hour}), and observables (\texttt{efforts}, \texttt{wkb}, \texttt{hrsncared}, \texttt{overwhelmed}). 
Exact bin edges, tolerance parameters, and the maximum history length $L_{\max}$ will be documented alongside aggregate results to facilitate verification and reuse.

\section*{Acknowledgments}
The simulator employed in this study was developed for research and experimental purposes in collaboration with municipal authorities. 
All data handling and processing comply with the General Data Protection Regulation (Regulation (EU) 2016/679) and the licensing conditions of the Italian National Institute of Statistics (ISTAT). 
The system is exempt from the provisions of the EU Artificial Intelligence Act (Regulation (EU) 2024/1689) under the research exemption and is not classified as medical-device software under Regulation (EU) 2017/745.

During manuscript preparation, the author used digital assistants (ChatGPT, OpenAI; Perplexity.ai, San Francisco, CA; and Publish or Perish, Harzing.com) solely for language refinement, LaTeX and code editing, literature retrieval, and bibliometric verification. 
The author reviewed and edited all generated content and assumes full responsibility for the final text and its scientific interpretations. 
No financial or material support, administrative assistance, or in-kind contributions were received beyond the software tools explicitly mentioned above.

\appendix
\section{Algorithmic Definitions and Implementation Details}
\label{app:algorithms}

This appendix provides the full algorithmic definitions referenced in
Sections~\ref{sec:framework}--\ref{sec:casestudy}. It collects reconstruction,
model-selection, and entropy-estimation details to avoid redundancy in the main
text.

\subsection*{Simulation Configuration}
\label{app:simulation-config}

Table~\ref{tab:simulationconfig} summarizes the core configuration parameters
of the agent-based simulation used to generate all time series analyzed in
this study. These parameters define the scale and temporal resolution of the
simulated system and are reported to facilitate replication and contextual
interpretation of the results. All analyses are conditional on this simulation
design.

\begin{table}[htbp]
	\centering
	\small
	\setlength{\tabcolsep}{6pt}
	\renewcommand{\arraystretch}{1.2}
	\caption{Core simulation configuration parameters.}
	\label{tab:simulationconfig}
	\begin{tabular}{p{5.2cm} p{6.8cm}}
		\toprule
		\textbf{Parameter} & \textbf{Value / Description} \\
		\midrule
		
		Simulation type & Agent-based model of caregiver--elder dyads \\
		Unit of analysis & Caregiver--elder dyad \\
		Number of dyads & Fixed per simulation run (order of $10^2$) \\
		Time step & Discrete simulation tick (uniform temporal resolution) \\
		Simulation horizon & Fixed number of ticks per run (order of $10^3$) \\
		Recorded variables &
		Caregiving effort (\textit{efforts}), walkability index (\textit{wkb}),
		hours not cared (\textit{hrsncared}), overwhelmed status (binary) \\
		Variable nature &
		Mixed continuous, zero-inflated, and binary variables \\
		Spatial structure &
		Explicit simulated physical environment with distances and service locations \\
		Initialization &
		Socioeconomic attributes fixed at initialization; behavioral and accessibility
		variables evolve over time \\
		Data origin &
		Fully simulated; no empirical or observational data used \\
		\bottomrule
	\end{tabular}
\end{table}

\subsection*{Compression-Based Complexity Proxies: Implementation Details}
\label{app:compression}

Compression-based measures are employed in this study as algorithm-agnostic
proxies for descriptive regularity and algorithmic complexity in symbolic
sequences generated by the agent-based model. These measures complement
$\varepsilon$-machine--based quantities by quantifying redundancy and
compressibility without making assumptions about predictive structure or
causal state representations.

All compression algorithms are applied using standard library implementations
with default parameter settings.\footnotemark{}
This choice is intentional: while parameter tuning can improve compression
performance for specific sequence lengths or symbol distributions, it also
introduces algorithm- and data-dependent bias. In a diagnostic and comparative
context---where the objective is to assess relative structure across variables,
dyads, and representations rather than to achieve optimal encoding---the use
of standardized defaults preserves methodological neutrality, reproducibility,
and comparability across algorithms.

For each symbolic sequence of length $n$, the compressed output size
$S_{\mathrm{comp}}$ (in bytes) is converted to a normalized measure of bits per
symbol,
\begin{equation}
	C_{\mathrm{comp}} = \frac{8\,S_{\mathrm{comp}}}{n},
\end{equation}
allowing direct comparison with entropy-rate estimates expressed in bits per
symbol. Compression is performed on raw byte streams representing the symbolic
sequences, without metadata or container-specific headers.

\begin{table}[htbp]
	\centering
	\small
	\setlength{\tabcolsep}{6pt}
	\renewcommand{\arraystretch}{1.2}
	\caption{Compression-based complexity proxies: algorithms and parameter settings.}
	\label{tab:compression-settings}
	\begin{tabular}{p{2.8cm} p{3.6cm} p{3.2cm} p{4.4cm}}
		\toprule
		\textbf{Algorithm} & \textbf{Implementation} & \textbf{Key parameters} & \textbf{Values used and rationale} \\
		\midrule
		
		\textbf{LZMA} (Lempel--Ziv--Markov) 
		& Python \texttt{lzma} (XZ Utils reference implementation) 
		& Compression preset; dictionary size; container format 
		& Default preset (\texttt{preset = 6}); dictionary size determined by preset (library default); \texttt{FORMAT\_XZ}. Defaults balance compression strength and computational cost and allow detection of long-range dependencies without algorithm-specific tuning. \\
		
		\textbf{BZ2} (Burrows--Wheeler transform) 
		& Python \texttt{bz2} 
		& Block size; entropy coding 
		& Default block size (\texttt{blocksize = 9}, 900~kB); Huffman coding enabled by default. Captures medium-range regularities with moderate computational cost. \\
		
		\textbf{GZIP} (DEFLATE: LZ77 + Huffman) 
		& Python \texttt{gzip} 
		& Compression level; sliding window 
		& Default maximum compression level (\texttt{compresslevel = 9}); sliding window fixed by DEFLATE standard (32~kB). Emphasizes short-range redundancy and fast execution. \\
		
		\textbf{All compressors} 
		& -- 
		& Input format; tuning policy; output metric 
		& Symbol sequences serialized as raw byte streams without metadata. No parameter tuning performed beyond library defaults to preserve comparability across variables and dyads. Compressed size $S_{\mathrm{comp}}$ normalized as bits per symbol: 
		$
		C_{\mathrm{comp}} = \frac{8\,S_{\mathrm{comp}}}{n},
		$
		where $n$ is the sequence length. \\
		
		\bottomrule
	\end{tabular}
\end{table}

\footnotetext{%
	Default compression settings are used intentionally. Parameter tuning can
	improve compression ratios for particular sequence lengths or symbol
	distributions, but it also introduces algorithm- and data-dependent bias. For
	diagnostic studies aimed at comparative analysis rather than optimal encoding,
	standardized defaults ensure neutrality, reproducibility, and consistency
	across compression algorithms.%
}

\subsection{System-Level $\varepsilon$-machine Reconstruction: Definitions}
In the context of agent-based models (ABMs), $\varepsilon$-machines provide a
principled means to extract and quantify a system’s intrinsic computation—the
transformation of past information into future behavior. Here, causal states
may correspond to recurrent behavioral regimes, coordination cycles, or
macro-level feedback patterns that emerge from agent interactions. Unlike
scalar measures such as entropy or mutual information alone, the
$\varepsilon$-machine reconstructs the \emph{minimal predictive model}: it
reveals the architecture of information storage and flow that governs the
system’s evolution. This makes it particularly suited to studying adaptive or
path-dependent ABM dynamics, where emergent regularities coexist with
stochastic fluctuations.

\medskip
The $\varepsilon$-machine provides the minimal unifilar model that partitions
past observation histories into sets—called causal states—that yield identical
conditional distributions over possible futures. Formally, if two pasts
$(x_{:t})$ and $(x'_{:t})$ satisfy
\begin{equation}
	\mathbb{P}\!\big(X_{t:\infty} \mid X_{:t}=x_{:t}\big)
	= \mathbb{P}\!\big(X_{t:\infty} \mid X_{:t}=x'_{:t}\big),
	\label{eq:causal_state_equiv}
\end{equation}
then they belong to the same causal state. The $\varepsilon$-machine thus
defines the minimal predictive architecture of the process, specifying how
information from the past is stored and used to generate the future.

\medskip
Three canonical quantities summarize its informational organization: the
entropy rate $h_{\mu}$ (average unpredictability per symbol), the statistical
complexity $C_{\mu}$ (information stored in the causal-state distribution), and
the excess entropy $E$ (shared information between past and future). In concise
form:
\begin{equation}
	h_{\mu} = H[X_t \mid X_{:t}],
	\label{eq:entropy_rate}
\end{equation}
\begin{equation}
	C_{\mu} = H[\mathcal{S}],
	\label{eq:statistical_complexity}
\end{equation}
\begin{equation}
	E = I[X_{:t};\, X_{t:\infty}],
	\label{eq:excess_entropy}
\end{equation}
where $H[\cdot]$ and $I[\cdot;\cdot]$ denote Shannon entropy and mutual
information, respectively, and $\mathcal{S}$ is the causal-state random
variable. These quantities capture how much information a process generates,
stores, and transmits over time, thereby distinguishing structured dynamics
from mere randomness.

\medskip
Accurate estimation of these quantities, however, critically depends on how
continuous simulation outputs are represented as discrete symbolic sequences.
Symbolization—the transformation of numeric or multidimensional trajectories
into finite alphabets—acts as the analytical lens through which temporal
organization becomes observable. An appropriate scheme determines which
variations are treated as equivalent and which are preserved as meaningful
distinctions. Poorly chosen symbolizations can obscure dependencies by merging
distinct states or, conversely, fabricate structure by discretizing noise into
artificial categories.

In agent-based simulations, this step is particularly delicate because
variables differ in scale and generative mechanism. Spatial indicators such as
\emph{walkability} exhibit smooth continuity and feedback, requiring ordinal or
rank-based encodings that preserve relative change, whereas socioeconomic
measures like \emph{caregiver effort} or \emph{hours not cared} vary primarily
across agents and are better discretized using quantile or hurdle-mixture
schemes. Symbolization thus defines the \emph{effective alphabet} and the
\emph{grain of observation}, balancing detail and parsimony. Finer encodings
may inflate entropy and complexity through over-partitioning, while coarser
ones suppress correlations and underestimate structure.

Beyond its technical role, symbolization is a modeling decision: it specifies
the mapping between the simulated micro-dynamics and the informational
architecture that $\varepsilon$-machines can reveal. From an information-theoretic perspective, this dependence on representation
is not a methodological flaw but a standard conditioning assumption. Entropy
and related quantities are defined relative to an observation alphabet, and
coarse-graining alters the measurable information content of a process without
changing its underlying generative mechanism. This is analogous to the role of
macrostates in statistical physics, where entropy depends on the chosen
partition of phase space rather than on microscopic dynamics
\cite{cover2006elements,shannon1948}.

\subsection{$\varepsilon$-Machine Reconstruction and Entropy Estimation Pipeline}
The pipeline transforms a univariate symbolic series into a minimal
predictive-state representation, evaluates both causal and block-entropy-based
complexity metrics, and produces graphical and tabular outputs for subsequent
analysis. Given an input sequence of discrete symbols
$\mathbf{s} = (s_1, s_2, \dots, s_n)$ with
$s_t \in \{0, \dots, A{-}1\}$, the alphabet size $A$ is inferred as
$A = \max(\mathbf{s}) + 1$ if not provided.

\subsubsection{$\varepsilon$-Machine Reconstruction}

For a given Markov order $L$, we denote by $s_{t-L:t-1}$ the length-$L$ history
of symbols immediately preceding time $t$.

The input sequence of integer symbols is first normalized, and the alphabet
size is inferred from the data unless explicitly specified. The main
reconstruction step builds a probabilistic, unifilar state model. Within this
stage, the optimal Markov order $L^{\ast}$ is determined by minimizing the
Bayesian Information Criterion (BIC) across candidate orders
$L = 0, \dots, L_{\max}$.

\medskip
Let $N(h,a)$ denote the number of occurrences of symbol
$a \in \{0, \dots, A{-}1\}$ following a history $h = s_{t-L:t-1}$, and let
$N(h) = \sum_{a} N(h,a)$ denote the total number of occurrences of history $h$.
For each candidate order $L$, the empirical next-symbol probabilities are then
defined as:
\begin{equation}
	\hat{P}(a \mid h) = \frac{N(h,a)}{N(h)} ,
	\label{eq:empirical_probabilities}
\end{equation}

The corresponding log-likelihood and Bayesian Information Criterion are given
by:
\begin{equation}
	\log \mathcal{L}_L =
	\sum_{h} \sum_{a} N(h,a)\,\log \hat{P}(a \mid h),
	\label{eq:log_likelihood}
\end{equation}
\begin{equation}
	\mathrm{BIC}(L) = -2 \log \mathcal{L}_L + k_L \log N_{\mathrm{eff}},
	\label{eq:bic_definition}
\end{equation}
where $k_L$ is the number of free parameters and
$N_{\mathrm{eff}} \approx n - 1 - L$. The order $L^{\ast}$ minimizing
$\mathrm{BIC}(L)$ is then selected as the optimal Markov order.

All subsequent reconstruction steps, including history clustering, transition
construction, and information-theoretic estimation, are performed using the
selected Markov order $L^{\ast}$ unless explicitly stated otherwise.

\subsubsection{Predictive Distributions and History Clustering}

For the selected order, predictive conditional distributions
$P(a_{t+1} \mid a_{t-L:t})$ are estimated for all histories observed with
sufficient frequency. Histories with statistically similar predictive
distributions (as measured by their $\ell_1$ norm) are clustered to form
equivalence classes representing causal states.

\medskip
For each observed history $h$, we estimate:
\begin{equation}
	\hat{P}(a \mid h) = \frac{N(h,a)}{N(h)}, \qquad a \in \{0, \dots, A{-}1\},
	\label{eq:conditional_probability}
\end{equation}
where $N(h,a)$ is the joint count of occurrences of history $h$ followed by
symbol $a$, and $N(h)$ is the total count of $h$. Histories observed fewer than
a threshold count are discarded to suppress noise. Two histories $h$ and $h'$
belong to the same causal state if their $\ell_1$ norm satisfies
\begin{equation}
	\big\| P(\cdot \mid h) - P(\cdot \mid h') \big\|_1 \le \mathrm{tol}.
	\label{eq:l1_distance}
\end{equation}

Each causal state $s$ thus represents a cluster of statistically equivalent
histories $C_s$. The mean emission distribution for each state is computed as
\begin{equation}
	\hat{P}_{\mathrm{emit}}(a \mid s) =
	\frac{1}{|C_s|} \sum_{h \in C_s} \hat{P}(a \mid h),
	\label{eq:mean_emission}
\end{equation}
defaulting to a uniform vector if no histories are present. The distribution $\hat{P}_{\mathrm{emit}}(\cdot \mid s)$ represents the empirical
symbol emission frequencies associated with causal state $s$ and does not
constitute a transition probability between states.

\subsubsection{Unifilar Transitions and Stationary Distribution}
A \emph{unifilar transition structure} is a state–transition topology used in
computational mechanics and predictive-state models (such as
$\varepsilon$-machines), where each state–symbol pair uniquely determines the
next state. In empirical reconstructions from finite data, unifilarity is
enforced approximately by constructing a deterministic transition function
via a majority rule over observed history shifts. The transition function $T(s_i,a)=s_j$ is deterministic: it specifies the unique next causal state reached from state $s_i$ upon emission of symbol $a$. Transition probabilities are induced indirectly via the emission distributions
$\hat{P}_{\mathrm{emit}}(\cdot \mid s_i)$ and the stationary distribution $\pi$,
and are not associated with $T$ itself.

\medskip
For each history $h \in C_s$ and symbol $a$, the shifted context is defined as
\begin{equation}
	h' = \mathrm{shift}_L(h,a),
	\label{eq:shift_function}
\end{equation}
where $\mathrm{shift}_L(\cdot,\cdot)$ appends symbol $a$ to $h$ and discards the
oldest symbol to preserve history length $L$. The next state $s'$ is then chosen
as the most frequently observed causal state reached by these shifted
histories. This procedure yields a deterministic mapping
$T(s,a) = s'$ for each state–symbol pair.

The corresponding state-transition matrix $M$ is defined as
\begin{equation}
	M_{ij} = \sum_{a} \hat{P}_{\mathrm{emit}}(a \mid s_i)\,
	\mathbf{1}\!\{T(s_i,a)=s_j\},
	\label{eq:transition_matrix}
\end{equation}
and is normalized row-wise to ensure stochasticity. The stationary distribution
$\pi$ is obtained as the fixed point of the Markov chain,
\begin{equation}
	\pi = \pi M,
	\qquad
	\pi^{(t+1)} = \pi^{(t)} M,
	\qquad
	\|\pi^{(t+1)} - \pi^{(t)}\|_1 < \mathrm{tol},
	\label{eq:stationary_distribution}
\end{equation}
where convergence is assessed using the $\ell_1$ norm.

States with similar emission distributions can optionally be merged using a
Jensen--Shannon divergence threshold, regularizing the model by removing
redundant predictive structures.

\subsubsection{Optional State Merging by Jensen--Shannon Divergence}

To regularize over-fragmented models, states with similar emission
distributions are merged when
\begin{equation}
	\mathrm{JS}\!\big(\hat{P}_{\mathrm{emit}}(\cdot \mid s),\,
	\hat{P}_{\mathrm{emit}}(\cdot \mid s')\big) \le \tau_{\mathrm{JS}},
	\label{eq:js_merge_condition}
\end{equation}
where $\mathrm{JS}(P,Q)$ denotes the Jensen--Shannon divergence and
$\tau_{\mathrm{JS}}$ controls the merging threshold. The divergence is defined as
\begin{equation}
	\mathrm{JS}(p,q)
	= \tfrac{1}{2} D_{\mathrm{KL}}(p \,\|\, m)
	+ \tfrac{1}{2} D_{\mathrm{KL}}(q \,\|\, m),
	\qquad
	m = \tfrac{1}{2}(p + q),
	\label{eq:js_definition}
\end{equation}
and after merging, the stationary distribution $\pi$ is recomputed.

\subsubsection{Information-Theoretic Quantities}

From the reconstructed $\varepsilon$-machine, the following
information-theoretic quantities are derived.

\paragraph{Entropy rate.}

\begin{equation}
	h_{\mu} = \sum_s \pi_s\,
	H\!\big(\hat{P}_{\mathrm{emit}}(\cdot \mid s)\big),
	\qquad
	H(p) = -\sum_i p_i \log_2 p_i,
	\label{eq:entropy_rate_machine}
\end{equation}

quantifying the average unpredictability of symbol generation under the
stationary distribution of causal states.

\paragraph{Statistical complexity.}

\begin{equation}
	C_{\mu} = H(\pi) = -\sum_s \pi_s \log_2 \pi_s,
	\label{eq:statistical_complexity_machine}
\end{equation}

measuring the amount of information stored in the stationary causal-state
distribution.

\paragraph{Predictive information proxy.}

Exact computation of the excess entropy $E = I[X_{:t}; X_{t:\infty}]$ is
generally intractable for finite reconstructed models. We therefore report a
one-step predictive information proxy based on the reconstructed
$\varepsilon$-machine. Let

\begin{equation}
	\hat{P}_{\mathrm{next}} =
	\sum_s \pi_s \hat{P}_{\mathrm{emit}}(\cdot \mid s),
	\qquad
	H_{\mathrm{next}} = H(\hat{P}_{\mathrm{next}}),
	\label{eq:predictive_proxy_pre}
\end{equation}

then

\begin{equation}
	E_{\mathrm{proxy}} =
	\max\!\big(H_{\mathrm{next}} - h_{\mu},\, 0\big),
	\label{eq:predictive_proxy}
\end{equation}

which approximates the amount of predictive structure captured at the
one-step horizon. By construction, $E_{\mathrm{proxy}}$ provides a lower-bound
proxy for the true excess entropy and is reported alongside block-based
estimates below.

Finite-block entropy estimates and excess-entropy approximations are reported
only as auxiliary diagnostics and are provided in Appendix~A for completeness.
They are not used for inference or model selection in the main analysis.

\subsubsection{Visualization and Data Export}

Finally, the pipeline generates the following visual and tabular artifacts:
\begin{enumerate}[label=(\roman*)]
	\item a time-series plot of the input symbols;
	\item a bar plot of stationary causal-state probabilities;
	\item a transition graph representing the reconstructed $\varepsilon$-machine;
	\item JSON and CSV exports describing states, emissions, and transitions; and
	\item a table of BIC scores across candidate Markov orders.
\end{enumerate}

The pipeline returns the selected Markov order $L^{\ast}$, the number of causal
states $|S|$, the entropy rate $h_{\mu}$, statistical complexity $C_{\mu}$, the
predictive information proxy $E_{\mathrm{proxy}}$, block-based estimates
$\big(h_{\mu}^{(\mathrm{block})}, E^{(\mathrm{block})}\big)$, and file paths for
all generated visual and tabular artifacts.

\subsection{Symbolization and Discretization Pipeline}

This pipeline provides a unified framework for transforming continuous variables
into discrete symbolic representations. It supports both \emph{data-driven}
(quantile, Jenks, Gaussian mixture, vector quantization) and \emph{rule-based}
(fixed, policy, winsorized) discretizations. The design is modular, with each
transformation stage isolated and equipped with fallback procedures to ensure
robustness under degenerate or ill-conditioned input data.

\medskip
Let the numeric series be denoted by
\begin{equation}
	\mathbf{x} = (x_1, \dots, x_n)^\top \in \mathbb{R}^n.
	\label{eq:numeric_series}
\end{equation}

A binning scheme with $k$ bins is defined by the ordered set of edges
\begin{equation}
	\mathbf{e} = (e_0, \dots, e_k),
	\qquad
	e_0 < \cdots < e_k.
	\label{eq:bin_edges}
\end{equation}

Each observation is assigned to a discrete label (symbol) according to
\begin{equation}
	\ell(x_i; \mathbf{e})
	= \min\!\big\{\, j \in \{0, \dots, k{-}1\} : e_j \le x_i < e_{j+1} \,\big\}.
	\label{eq:label_assignment}
\end{equation}

The alphabet size is $A = k$ (or fewer if bins are empty). Unless otherwise
specified, bins are left-closed and right-open, $[e_j, e_{j+1})$. Ties and
non-finite values are handled by small perturbations or exclusion before
computing quantiles or class boundaries.

\subsubsection{Discretization Methods}

Each of the methods described below produces a label vector
$\boldsymbol{\ell}$, an edge set $\mathbf{e}$ (if applicable), and the alphabet
size $A$. All methods follow the same high-level transformation sequence:
\begin{center}
	\ttfamily
	Input series $\rightarrow$ [optional winsorization] $\rightarrow$
	edge computation $\rightarrow$ digitization $\rightarrow$
	symbolic labeling $\rightarrow$
	output (labels, edges, alphabet size)
\end{center}

Here, the parameter $m$ denotes the local embedding or pattern order used for
symbolization and is conceptually distinct from the Markov order $L$ employed
later in $\varepsilon$-machine reconstruction.

Fallback mechanisms ensure stability: Jenks $\rightarrow$ K-Means $\rightarrow$
quantiles for class breaks, and default clipping to small perturbations when
numerical degeneracy occurs. These fallback procedures are purely algorithmic
safeguards to ensure numerical stability and do not introduce additional
modeling assumptions.

\paragraph{1. Quantile Bins.}
Let $Q(p)$ denote the empirical quantile of $\{x_i\}$ at cumulative probability
$p$. Define bin edges as
\begin{equation}
	e_j = Q\!\left(\frac{j}{k}\right),
	\qquad j = 0, \dots, k.
	\label{eq:quantile_bins}
\end{equation}
After enforcing monotonicity, assign labels $\ell_i = \ell(x_i; \mathbf{e})$.
Alphabet size: $A = k$.

\paragraph{2. Fixed or Policy-Fixed Edges.}
Given explicit user-supplied cutpoints $e_0 < e_1 < \dots < e_k$, assign labels
via $\ell_i = \ell(x_i; \mathbf{e})$. Alphabet size: $A = k$.

\paragraph{3. Winsorized Equal-Width.}
Using parameters $p_{\mathrm{low}}$ and $p_{\mathrm{high}}$, compute
\begin{equation}
	\ell = Q(p_{\mathrm{low}}),
	\qquad
	h = Q(p_{\mathrm{high}}),
	\label{eq:winsor_limits}
\end{equation}
then clip as in Section~\ref{sec:winsorization}. Edges are constructed as
\begin{equation}
	e_j = \ell + \frac{j}{k}(h - \ell),
	\qquad j = 0, \dots, k.
	\label{eq:winsor_edges}
\end{equation}
Alphabet size: $A = k$.

\paragraph{4. Jenks Natural Breaks.}

Use edges $\mathbf{e}^\star$ obtained from the Jenks optimization in Section~\ref{sec:jenks}. 
Assign labels $\ell_i = \ell(x_i; \mathbf{e}^\star)$ and set alphabet size $A = k$.

\paragraph{5. Hurdle + Gaussian Mixture Model.}

Nonpositive values form a separate bin ($\ell_i = 0$ if $x_i \le 0$). 
For the positive subset $X^+ = \{x_i > 0\}$, fit Gaussian mixtures of varying component count $m$:

\begin{equation}
	f(x; \theta_m) = \sum_{c=1}^{m} \pi_c \, \mathcal{N}(x \mid \mu_c, \sigma_c^2).
	\label{eq:gmm_definition}
\end{equation}

Select the optimal $\hat{m}$ by minimizing the Bayesian Information Criterion (BIC; \cite{schwarz1978}):

\begin{equation}
	\mathrm{BIC}(m) = -2 \sum_{x \in X^+} \log f(x; \theta_m) + p_m \log |X^+|.
	\label{eq:bic_selection}
\end{equation}

Assign each positive value to its maximum-posterior component index,

\begin{equation}
	\tilde{\ell}(x) = \arg\max_c \pi_c \, \mathcal{N}(x \mid \mu_c, \sigma_c^2),
	\label{eq:gmm_assignment}
\end{equation}

and shift labels by one to preserve the zero bin:

\begin{equation}
	\ell(x) = 
	\begin{cases}
		0, & x \le 0, \\[4pt]
		\tilde{\ell}(x) + 1, & x > 0.
	\end{cases}
	\label{eq:gmm_labels}
\end{equation}

Alphabet size: $A = \hat{m} + 1$.

\paragraph{6. Ordinal Patterns.}

For an order parameter $m$, construct sliding windows

\begin{equation}
	\mathbf{w}_t = (x_{t-m+1}, \dots, x_t).
	\label{eq:ordinal_windows}
\end{equation}

Let $\pi_t$ be the permutation that sorts $\mathbf{w}_t$ increasingly. 
Encode $\pi_t$ into an integer index using the Lehmer code:

\begin{equation}
	c_{t,i} = \#\{\, j > i : \pi_t(j) < \pi_t(i) \,\}, 
	\qquad
	\mathrm{idx}(\pi_t) = \sum_{i=1}^{m} c_{t,i} (m - i)!.
	\label{eq:lehmer_code}
\end{equation}

Set $\ell_t = \mathrm{idx}(\pi_t)$ and pad the first $(m{-}1)$ indices. 
Alphabet size: $A = m!$.

\paragraph{7. Lagged Vector Quantization.}

For lag order $m$, build lagged vectors

\begin{equation}
	\mathbf{y}_t = (x_{t-m+1}, \dots, x_t)^\top, 
	\qquad t = m, \dots, n.
	\label{eq:lag_vectors}
\end{equation}

Optionally standardize as

\begin{equation}
	\mathbf{z}_t = (\mathbf{y}_t - \hat{\mu}) / \hat{\sigma},
	\label{eq:standardization}
\end{equation}

(elementwise). Apply $k_{\mathrm{VQ}}$-means clustering to minimize

\begin{equation}
	\sum_{t=m}^{n} \min_j \|\mathbf{z}_t - \mathbf{c}_j\|_2^2.
	\label{eq:vq_optimization}
\end{equation}

Assign each vector to its nearest centroid:

\begin{equation}
	\ell_t = \arg\min_j \|\mathbf{z}_t - \mathbf{c}_j\|_2^2.
	\label{eq:vq_assignment}
\end{equation}

Pad the first $(m{-}1)$ samples with $\ell_m$. 
Alphabet size: $A = k_{\mathrm{VQ}}$.

\subsubsection{Binning, Digitization, and Label Formation}

Discrete labels are assigned from predefined boundaries (digitization), generating bin edges through equal-width or Jenks optimization, and optionally preprocessing data by clipping extreme values (winsorization).

\paragraph{Digitization.}

Given edges $\mathbf{e}$, integer labels are computed as

\begin{equation}
	\ell_i = \sum_{j=1}^{k-1} \mathbb{I}\{x_i \ge e_j\}, 
	\qquad i = 1, \dots, n.
	\label{eq:digitization}
\end{equation}


\paragraph{Equal-Width Edges.}

For given lower and upper endpoints 
$\ell = \min_i x_i$ and $h = \max_i x_i$ (optionally provided), 
define $k$ equal-width intervals with edges

\begin{equation}
	e_j = \ell + \frac{j}{k}(h - \ell), 
	\qquad j = 0, \dots, k.
	\label{eq:equalwidth_edges}
\end{equation}

\paragraph{Jenks Natural Breaks.}
\label{sec:jenks}
Jenks natural breaks~\cite{jenks1967} minimize the within-class sum of squares:

\begin{equation}
	\mathcal{J}(\mathbf{e}) = 
	\sum_{c=1}^k \sum_{x_i \in C_c} (x_i - \bar{x}_{C_c})^2,
	\label{eq:jenks_objective}
\end{equation}

where $\bar{x}_{C_c}$ is the class mean. 
The objective is equivalent to maximizing the goodness-of-variance-fit (GVF):

\begin{equation}
	\mathrm{GVF} = 
	1 - 
	\frac{\sum_{c=1}^k \sum_{x_i \in C_c} (x_i - \bar{x}_{C_c})^2}{
		\sum_{i=1}^n (x_i - \bar{x})^2},
	\qquad
	\bar{x} = \frac{1}{n}\sum_i x_i.
	\label{eq:jenks_gvf}
\end{equation}


\paragraph{Winsorization.}
\label{sec:winsorization}
Given probabilities $p_{\text{low}}$ and $p_{\text{high}}$, define the lower and upper quantile thresholds as

\begin{equation}
	\ell = Q(p_{\text{low}}),
	\qquad
	h = Q(p_{\text{high}}),
	\label{eq:winsor_thresholds}
\end{equation}

where $Q(p)$ is the empirical quantile function of $\mathbf{x}$. 
Each value is clipped to this interval as

\begin{equation}
	x_i^{\text{win}} = \min\{ \max\{x_i, \ell\}, \, h \}.
	\label{eq:winsor_clipping}
\end{equation}

The function returns $(\mathbf{x}^{\text{win}}, \ell, h)$.

\section{Finite-Block Entropy and Excess Entropy Diagnostics}
This appendix reports finite-block entropy estimates as a diagnostic consistency
check against $\varepsilon$-machine–based quantities.

An independent estimate of the entropy rate and excess entropy is obtained via
finite-block entropies up to block length $L_{\max}$. For each block length $L$,
the empirical block entropy $H_L$ is estimated using a plug-in estimator with
mild additive smoothing. Asymptotic values are approximated as

\begin{equation}
	h_{\mu}^{(\mathrm{block})} \approx H_L - H_{L-1},
	\qquad
	E^{(\mathrm{block})} \approx H_L - L\,h_{\mu}^{(\mathrm{block})},
	\label{eq:block_extrapolation}
\end{equation}

which follow standard finite-difference approximations for stationary symbolic
processes.

For observed $L$-blocks $b \in \mathcal{B}^L$ with counts $N(b)$ and total
$N = \sum_b N(b)$, smoothed probabilities are estimated as

\begin{equation}
	\hat{P}(b) = \frac{N(b) + 1}{N + |\mathcal{B}^L|},
	\label{eq:smoothed_prob}
\end{equation}

and the corresponding block entropy is computed as

\begin{equation}
	H_L = -\sum_b \hat{P}(b)\log_2 \hat{P}(b).
	\label{eq:block_entropy}
\end{equation}

To reduce finite-sample noise, block-based estimates are refined by averaging
the final three increments of $H_L - H_{L-1}$:

\begin{equation}
	h_{\mu}^{(\mathrm{block})}
	= \mathrm{mean}\!\big[H_L - H_{L-1}\big]_{L_{\max}-2}^{L_{\max}},
	\qquad
	E^{(\mathrm{block})}
	= \max\!\big(H_{L_{\max}} - L_{\max} h_{\mu}^{(\mathrm{block})},\, 0\big).
	\label{eq:block_refined}
\end{equation}




\end{document}